\documentclass[pre,twocolumn,superscriptaddress,showpacs,pdfoutput=1]{revtex4-1}
\usepackage{graphicx}
\usepackage{epsfig}
\usepackage{amssymb,amsmath,amsfonts,hyperref}
\usepackage{wasysym}
\usepackage{bm}
\usepackage{latexsym}
\usepackage{eucal}
\usepackage{float}
\usepackage{verbatim}
\usepackage[normalem]{ulem}
\usepackage{epstopdf}
\usepackage{color}
\usepackage{braket}

\newcommand{\be}{\begin{equation}}
\newcommand{\ee}{\end{equation}}

\newcommand{\eea}{\end{eqnarray}}

\newcommand{\xp}{{x^{\prime}}}
\newcommand{\yp}{{y^{\prime}}}
\newcommand{\xpp}{{x^{\prime\prime}}}
\newcommand{\ypp}{{y^{\prime\prime}}}
\newcommand{\tp}{{t^{\prime}}}
\newcommand{\pt}{{\partial_t}}
\newcommand{\px}{{\partial_x}}
\newcommand{\pxx}{{\partial_x^2}}
\newcommand{\py}{{\partial_y}}
\newcommand{\pyy}{{\partial_y^2}}
\newcommand{\prmx}{{x^{\prime}}}
\newcommand{\prmy}{{y^{\prime}}}

\begin{document}

\title{Anomalous relaxation of density waves in a
ring-exchange system}

\author{Pranay Patil}
\affiliation{Max-Planck Institute for the Physics of Complex Systems,
Dresden, Germany}
\affiliation{Laboratoire de Physique Th\'eorique, Universit\'e de Toulouse, CNRS, UPS, France}

\author{Markus Heyl}
\affiliation{Max-Planck Institute for the Physics of Complex Systems,
Dresden, Germany}
\affiliation{Theoretical Physics III, Center for Electronic Correlations and Magnetism, Institute of Physics, University of Augsburg, 86135 Augsburg, Germany}

\author{Fabien Alet}
\affiliation{Laboratoire de Physique Th\'eorique, Universit\'e de Toulouse, CNRS, UPS, France}

\begin{abstract}
We present the analysis of the slowing down exhibited by stochastic dynamics of a ring-exchange model on a square lattice, by means of numerical simulations. We find the preservation of coarse-grained memory of initial state of density-wave types for unexpectedly long times. This behavior is inconsistent with the prediction from a low frequency continuum theory developed by assuming a mean field
solution. Through a detailed analysis of correlation functions of the dynamically active regions, we exhibit an 
 unconventional transient long ranged structure formation in a direction which is featureless
for the initial condition, and argue that its slow melting plays a crucial role in the slowing-down mechanism.
We expect our results to be relevant also for the dynamics of quantum ring-exchange
dynamics of hard-core bosons and more generally for dipole
moment conserving models.
\end{abstract}

\maketitle

\section{Introduction}

The field of dynamics in isolated quantum systems has recently received
an increasing amount of attention thanks to the discoveries of a plethora of
interesting non-equilibrium behaviors \cite{calabrese2018entanglement,garrahan2018aspects,gogolin2016equilibration},
and of versatile experimental platforms to realize the same
\cite{browaeys2020many,gross2017quantum}.
These studies have been partially motivated by the desire to achieve the
protection of quantum information from scrambling caused by Hamiltonian
dynamics or environmental noise. This has lead to the rapid development
of the field of many body localization \cite{abanin2019,alet2018many,abanin2017recent,luitz2015many},
which relies on strong disorder to provide 
a safeguard against scrambling many body
dynamics, and the nascent field of Hilbert space fragmentation
\cite{sala2020ergodicity,khemani2020localization,moudgalya2021quantum,yang2020hilbert,mukherjee2021minimal,karpov2021disorder},
which results from the highly constrained configuration space of
Hamiltonians with a large number of strong local constraints and/or highly
frustrated interactions \cite{zhang2022hilbert,sikora2011extended}. 

A milder version of the total arresting of dynamics generated by the
phenomena mentioned above is realized by systems which approach
equilibrium in a manner which is qualitatively different from standard
diffusion. An example of this has recently been explored
for the spin-1/2 Heisenberg chain as well as its classical version
at intermediate energies
where evidence of super-diffusion has recently been seen both
theoretically \cite{ilievski2018superdiffusion,mcroberts2022anomalous}
and experimentally \cite{scheie2021detection},
leading to connections to surface growth
dynamics studied by Kardar-Parisi-Zhang \cite{weiner2020high,kardar1986dynamic}.
An approach to equilibrium which is slower than that expected from
diffusion has also been realized in systems which conserve higher
moments (such as dipolar and octupolar) of the spin
configuration \cite{iaconis2019anomalous,RefFP,morningstar2020kinetically,iaconis2021multipole,richter2022anomalous}.
For two- and higher-dimensional systems, this has also lead to the
realization of exotic fractonic
phases \cite{gromov2020fracton,nandkishore2019fractons,pretko2020fracton}.
Remarkable advances have also been made on probing experimental realizations of Hilbert space fragmentation and/or higher-moment conservation and associated sub-diffusion~\cite{guardado_2020,scherg_observing_2021,kohlert_2021}. 

While progress has been made on the analytical description and experimental detection of
the phenomena mentioned above, there has been a dire need for numerical
simulations on microscopic models to lend support to many of the predictions. This is in general
a difficult task as powerful methods to simulate large-scale quantum dynamics on the needed long-time scales are relatively few and capable of handling only specific regimes
\cite{baiardi2019large,schmitt2020quantum}.
It was found insightful~\cite{iaconis2019anomalous,iadecola2020nonergodic} to adopt methods from stochastic classical
mechanics  in the framework of cellular automata~\cite{iaconis2019anomalous,gopalakrishnan_2018,iaconis_prx_2021}, 
which ignore part of the quantum phase fluctuations and have been able to successfully
describe the long-term dynamical behavior of strongly interacting quantum systems.
This intuition arises from the expectation that for
generic systems with sufficiently large Hilbert
spaces and for times long compared to the microscopic energy scales of the Hamiltonian, the dynamics does
not show quantum coherence, thus reducing to a classical dynamics problem.
As mentioned above, well-known exceptions to this exist, but as
simulation of exact quantum dynamics is out of reach using current
methods, a study of the classical equivalent becomes of particular interest.
This can also serve as a natural starting point to understand the
complete quantum dynamics.
The language of cellular automata also lends itself naturally to
a hydrodynamic treatment, which identifies
an equivalence between quantum many-body dynamics at late times
and classical transport of globally conserved quantities
\cite{nahum2017,gopalakrishnan2018operator,iaconis_prx_2021}.
Slow thermalization
is often expected in integrable systems described by generalized
hydrodynamics \cite{castro2016emergent,bertini2016transport}. 
Quantum equivalents of cellular automata, which may be expected to
capture the dynamics more accurately, have also been found to
share important characteristics of integrable systems
\cite{klobas2021exact,alba2019operator}.

Following up on the studies of constrained systems, we consider in this work the
case of a simple hard-core bosonic model living on a square lattice,
undergoing ring-exchange dynamics. This model has already been
studied from the perspective of cuprates, as they serve as promising
candidates to realize high-Tc superconductivity \cite{paramekanti2002ring,rousseau2005ring,schaffer2009superfluid}.
Although traditionally most studies have focused on the possible
exotic ground state features of this model
\cite{tay2010possible,tay2011possible},
some recent works \cite{iaconis2019anomalous,RefC}
have considered the constrained dynamics generated by the ring-exchange mechanism, including starting from random configurations~\cite{iaconis2019anomalous}.
However, the
relevance of fragmentation to generic low-momentum states which are only described by
macroscopic patterns has not been investigated.

In this work we address this question using a classical approach based on stochastic dynamics. We find that structured initial configurations
in the form of a boson density wave retain their coarse grained structure
for a time which grows as a tunable power of the
wavelength, with an eventual melting which
is approximately described by a continuum model derived from
a simple Taylor expansion.
We study the detailed structure of the melting process via spatial
correlation functions and find that the dynamics proceed through the
development of strongly correlated large active regions which merge and
destroy the initial modulated pattern.

The detailed plan of the paper is as follows.
In Sec.\ref{S2}, we present the model, discuss the quantities conserved
under the dynamics, and elucidate the general profile of stripe-like 
configurations
which are perfectly frozen under the dynamics. The bulk of this section
discusses the effects of small perturbations on these exactly frozen
patterns, and the preservation of the memory of the initial state to
infinite times as illustrated by simulation of exact quantum and
stochastic classical dynamics on a small system size.
The close agreement between exact quantum dynamics and its stochastic
classical equivalent seen in this section motivates our approximation
and we focus purely on the classical system for the following sections.

Sec.\ref{S3} recalls the expected continuum field theory based on Taylor
expansions by first considering
the simpler case of correlated random walkers,
along with numerical checks of the equations developed. This is
followed by a treatment of the hard-core model using a
mean-field assumption.

For Sec.\ref{S4}, we move to more general configurations which take the
form of boson density waves, and show the persistence of the memory of
the initial state for unexpectedly long periods of time. We also compare
the prediction of the continuum field theory developed in Sec.\ref{S3}
with the initial dynamics. 

We follow up in Sec.\ref{S5} with a detailed analysis of the evolution of
the dynamical active regions, and present a phenomenological picture of
the mechanism leading up to the melting of the initial density wave
configuration. 

We summarize our results in Sec.\ref{S6} and present possibilities for
future follow ups via direct simulation of the quantum many body
dynamics.

In a following appendix \ref{AA},
we briefly discuss the long-time momentum space
profile of correlation functions, show that it is consistent
with a mean-field treatment obtained in a previous work, and that it cannot be a basis of the anomalous scaling observed in the present work (as could have been deduced from a recent analysis~\cite{sala2021dynamics} of a related model).

\section{Ring-exchange boson dynamics, conserved quantities and
frozen patterns}\label{S2}

\begin{figure}[t]
\includegraphics[width=\linewidth]{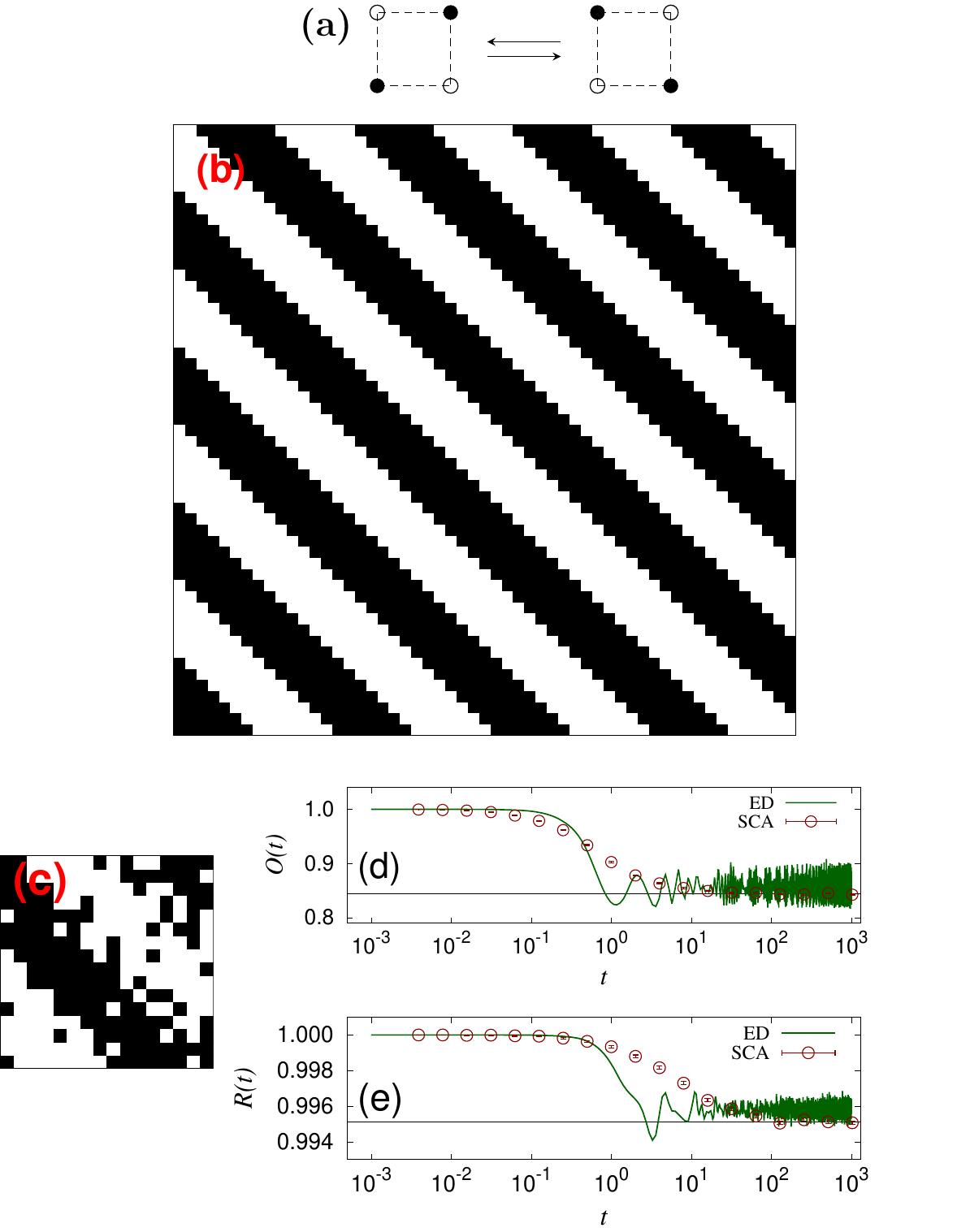}
\caption{
(a) The ring-exchange rule that is the only allowed dynamics in our system.
(b) A perfect stripe configuration which is frozen under the dynamic
rule above.
(c) Initial configuration with mixed filled-staggered pattern on a $L=16$ sample used to seed Hilbert
space fragment. The dynamics of the (d) overlap $O(t)$
and (e) Fourier ratio $R(t)$
with this initial state
for both exact dynamics (ED) in the quantum case and stochastic classical automaton
(SCA) dynamics, which quickly saturate to the exact
values calculated from enumerating all connected configurations.}
\label{figdef}
\end{figure}

We consider a system of hard-core bosons living on a square lattice,
which evolve stochastically in time using only ring-exchange dynamics
where bosons hop by pairs around a 4-sites plaquette of the lattice if and only if a single diagonal of the plaquette is occupied by two bosons (``flippable'' plaquettes). This only allowed dynamics is shown in Fig.~\ref{figdef}. This dynamic
rule trivially conserves the total particle number as well as the number of
particles in each individual column and in each individual row
\cite{tay2010possible,tay2011possible,iaconis2019anomalous,RefC}. 
For the rest of this work, we restrict ourselves to half-filling,
i.e. $L^2/2$ sites occupied by bosons, on a periodic $L\times L$ lattice,
and to the
sector where each column and each row has exactly $L/2$ occupied sites. 
One expects it to be the sector with the largest number
of configurations, as it is maximally symmetric. The total number of configurations in this sector can be computed on large lattices using combinatorial techniques~\cite{canfield2005asymptotic}.

\begin{figure}[t]
\includegraphics[width=\linewidth]{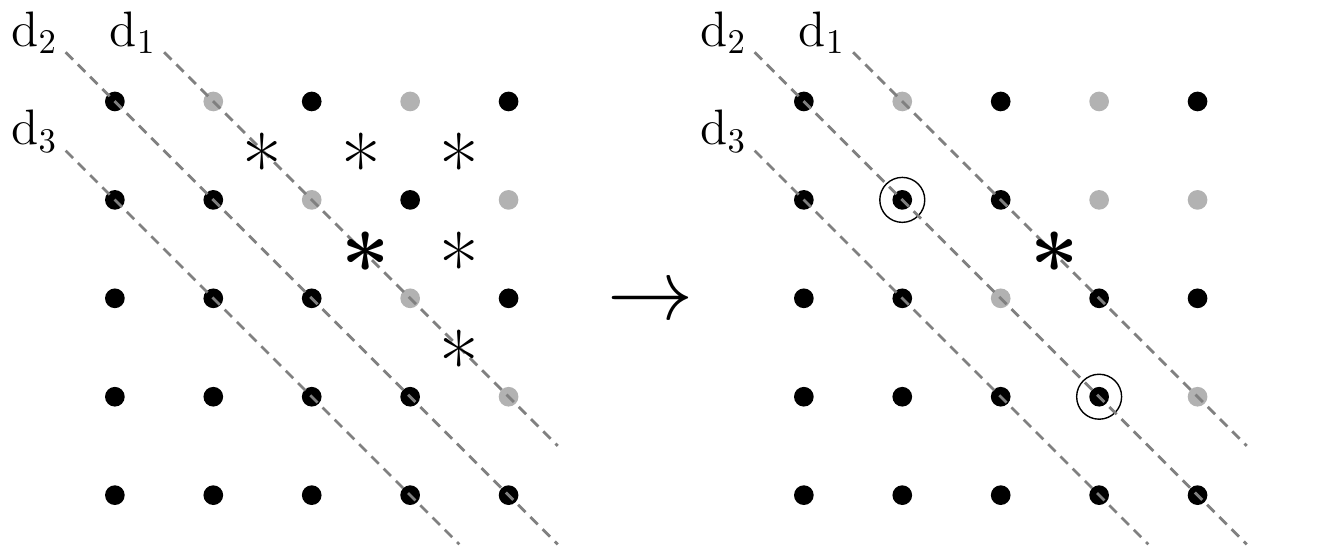}
\caption{Illustration of the melting of a N\'eel-filled boundary: Asterisks
mark flippable plaquettes, and the transformation from left to right takes
place by an exchange around the plaquette marked by the asterisk in bold type.
d$_i$ mark diagonals around the domain boundary.}
\label{figcartoon}
\end{figure}

We find that the conserved quantities discussed above do not in themselves
completely describe the dynamically connected sets of configurations. This
can be seen by considering a ``perfect'' stripe configuration as shown in
Fig.~\ref{figdef}, where an alternating pattern of filled and empty sites
does not leave a single flippable plaquette, making the configuration
frozen under ring-exchange dynamics.
By varying the widths and locations
of the filled and empty stripes, we can create many similar frozen
configurations. One may also naively expect that this restriction on
the number of accessible configurations extends to the case where we do not
have perfect stripes. To see this, we consider a configuration generated
by interchanging the two neighboring diagonals on the edge between a
filled and an empty region. This creates a diagonal made of flippable
plaquettes, where the influence of this active region may be expected to
only extend to a few lattice spacings around this diagonal.
Note that similar arguments are expected to hold more general
dynamics which conserve the dipole moment. To understand this
intuitively, we can consider larger plaquette dynamics, for example
2$\times$1 or 2$\times$2 plaquettes; once again the perfect stripe
configuration is frozen and spatial perturbations around it can
be expected to at most soften the boundary dynamically,
but still leave the stripes intact provided they are wide enough.
A simple example of the effect of the 1$\times$1 plaquette dynamics
on a boundary between a N\'eel like (highly active) and a
fully filled (inactive) region is shown in Fig.~\ref{figcartoon}.
Here one can see that although it is easy to
transfer a hole from diagonal $d_1$ to $d_2$, doing the same for $d_3$
would require holes at the circled locations, and thus
it is not possible to propagate our excitation
into the inactive region without sourcing another flippable
plaquette from within the active region.

To test this intuition and illustrate this effect, we exactly enumerate all the configurations connected
to an initial condition of the form described above
(slight random perturbation to the perfect stripe configuration)
for a $16\times 16$
lattice (shown in Fig.~\ref{figdef}).
We find that the total number of configurations,
which we call $N_c$
which belong to this ``fragment''
is $27,990$. To show that they retain some structure of the representative which
we have chosen, we compute an overlap as defined below,
\be
\bar{O}=\frac{4}{N_cL^2}\sum_c \sum_{x,y} n^{\rm seed}_{x,y} n^{c}_{x,y}.
\ee
where $n_{x,y}^{c / {\rm seed}}=0,1$ is the number of bosons at site of coordinates $(x,y)$ in the ($c$-th/initial) configuration. If we use all possible configurations (in which case $N_c$ would be the total
Hilbert space size), we would expect $\bar{O}=0$ due to the symmetry under
$n\to 1-n$. For the restricted Hilbert space belonging to the fragment
being considered we find $\bar{0}=0.844003...$, showing that for a dynamic
simulation restricted to this sector, the initial and late time states would
still retain significant overlap. To confirm this, we also run a stochastic
classical automaton (SCA) simulation, where at each time step, we propose $L^2$ random plaquette
flips, and if the chosen plaquette is flippable, we flip it with probability
$1/2$. The resulting overlap with the initial state is shown in
Fig.~\ref{figdef}, and we see that it quickly approaches the value
expected from exact enumeration, and retains it indefinitely.
To study the accuracy with which the SCA reproduces the exact quantum dynamics, we also
perform an exact diagonalization for this Hilbert space fragment, and compute
$\braket{\psi(t)|O|\psi(t)}$ for $\ket{\psi(t)}=e^{-iHt}\ket{\psi_0}$
using $H=\sum_{x,y}b^{\dagger}_{x,y}b_{x+1,y}b_{x,y+1}b^{\dagger}_{x+1,y+1}
+h.c.$,
where the initial condition is the
same occupation-basis state as we initialized the SCA with and $O$ now denotes an operator
which is diagonal in the occupation-basis and measures the overlap with the initial state.
This operator can be generated directly from the expression above for the classical case by
promoting $n^{c}_{x,y}$ to the number operator,
while retaining the integer status of $n^{\rm seed}_{x,y}$.
As shown in Fig.~\ref{figdef}, we find that the
overlap in the quantum dynamics closely traces the SCA during the initial decay away from
an overlap of $1$.

In the following sections, we quantify the relaxation to
equilibrium using the Fourier transform $n_k(t)$ with
frequency $k$ along the $\hat{x}+\hat{y}$ direction,
at time $t$. To this end, we measure the Fourier
ratio $R(t)=n_k(t)/n_k(0)$, and make a similar comparison
as done for the spatial overlap above. This is presented
in Fig.~\ref{figdef} as well, and once again we find a
close agreement between the exact quantum dynamics
and it's classical equivalent in the way they approach the
equilibrium value of $R(t)$ within the sector of interest.
This result suggests that the quantum dynamics within
sectors matches the classical automaton upon coarse graining
past time scales of $O$(1), suggesting
that quantum phases do not play a substantial
role in the aspects of dynamics which we want to study,
and that a classical
automaton approach could be sufficient to study the effect
of the kinetic constraints on large scale features.

Following the arguments above and noting that
exact quantum dynamics for larger sizes is not possible with current computational capabilities,
we directly study the long time behavior of our
stochastic dynamics simulations to gather information about
large system sizes in the following sections.

\section{Expected hydrodynamic description}\label{S3}

Here we discuss the coarse grained description in continuous space,
of the microscopic dynamics which we have introduced in the previous section. We begin by relaxing the hard-core constraint and considering the limit of
average number of particles per site being $\gg 1$. This allows us to
reduce the problem to non-interacting correlated random walkers, and leads to
an analogue of the diffusion equation which encodes vanilla sub-diffusion.
We show numerical evidence for the validity of the same.
The expected continuum dynamical behavior of this type of ring-exchange in 2d was first presented in Ref.~\cite{iaconis2019anomalous}, but we recall it for completeness as well as to understand how it should be perturbed to take into account the hard-core nature.
Next we move to an equivalent model of a real-valued field on the lattice
which makes the connection to the hydrodynamic limit more transparent,
and we again provide support with numerical simulations.

Lastly, we return to the hard-core model presented in the previous section,
and show that the continuum theory describing the model must include non-linear
terms in addition to the sub-diffusive term, and we show the possibility for
a quantitative change in the behavior of the system due to the non-linearities.

\subsection{Large particle number limit}

We relax the hard-core constraint of the stripe configurations described in the
previous section and assume the pattern to exist over a featureless
background of an average density of $n_d \gg1$ particles per site. The dynamics can now be
understood in terms of correlated random walkers in the following way.
First, we label each particle in the system as an independent walker.
A move is defined now as first picking a walker (called walker $a$)
at random, picking one of its
four next nearest neighboring sites with probability $1/4$, and moving one
of the walkers on the chosen site (called walker $b$) in tandem with
walker $a$ in a ring exchange type manner.
Due to the large number of walkers per site, we expect to always be able to
find walker $b$.

To write down the number of particles $n$ at a
particular position $(x,y)$ at time $t+1$ as a function of the values at
time $t$, we must consider all processes which can change $n_{x,y}$.
These processes are (1) choosing a walker at $(x,y)$ and moving it away
($\propto n_{x,y}$);
(2) choosing a walker at one of the four nearest neighbor sites and moving
it or its partner to $(x,y)$ ($\propto 1/2(n_{x\pm 1,y})$ or
$1/2(n_{x,y\pm 1})$);
(3) choosing one of the walkers at one of the next nearest
neighbors and moving it in tandem with a walker at $(x,y)$, thus reducing
the number of walkers at $(x,y)$ by one
($\propto 1/4(n_{x\pm 1,y\pm 1})$). Putting these terms together, we
can write the change in $n_{x,y}$, which is an whole number, from time
$t$ to time $t+1$ as (note that all terms in the expression below are at
time $t$, and we do not explicitly mention it for ease of representation)

\begin{align}
\Delta_t  & n_{x,y} \propto -n_{x,y} \nonumber\\
&+\frac{1}{2}[n_{x+1,y}+n_{x-1,y}+n_{x,y+1}+n_{x,y-1}] \nonumber \\
&-\frac{1}{4}[n_{x+1,y+1}+n_{x+1,y-1}+n_{x-1,y+1}+n_{x-1,y-1}].
\label{eq:dyn}
\end{align}

This equation was obtained using similar arguments in Ref.~\cite{iaconis2019anomalous}. 
If the system is initialized over a background of $n_d$ particles per site
using the stripe configuration described in the previous section (shown in
Fig.~\ref{figdef}), $n_{(x,y)}$
can be seen as a step function switching periodically between $n_d$ and
$n_d+1$. We would naively expect that the correlated dynamics discussed above
would quickly eliminate the sharp boundaries of the $n_{(x,y)}$ texture
and lead to a smooth function once we
average over many realizations of the stochastic dynamics. 

\begin{figure}[t]
\begin{center}
\includegraphics[width=\linewidth]{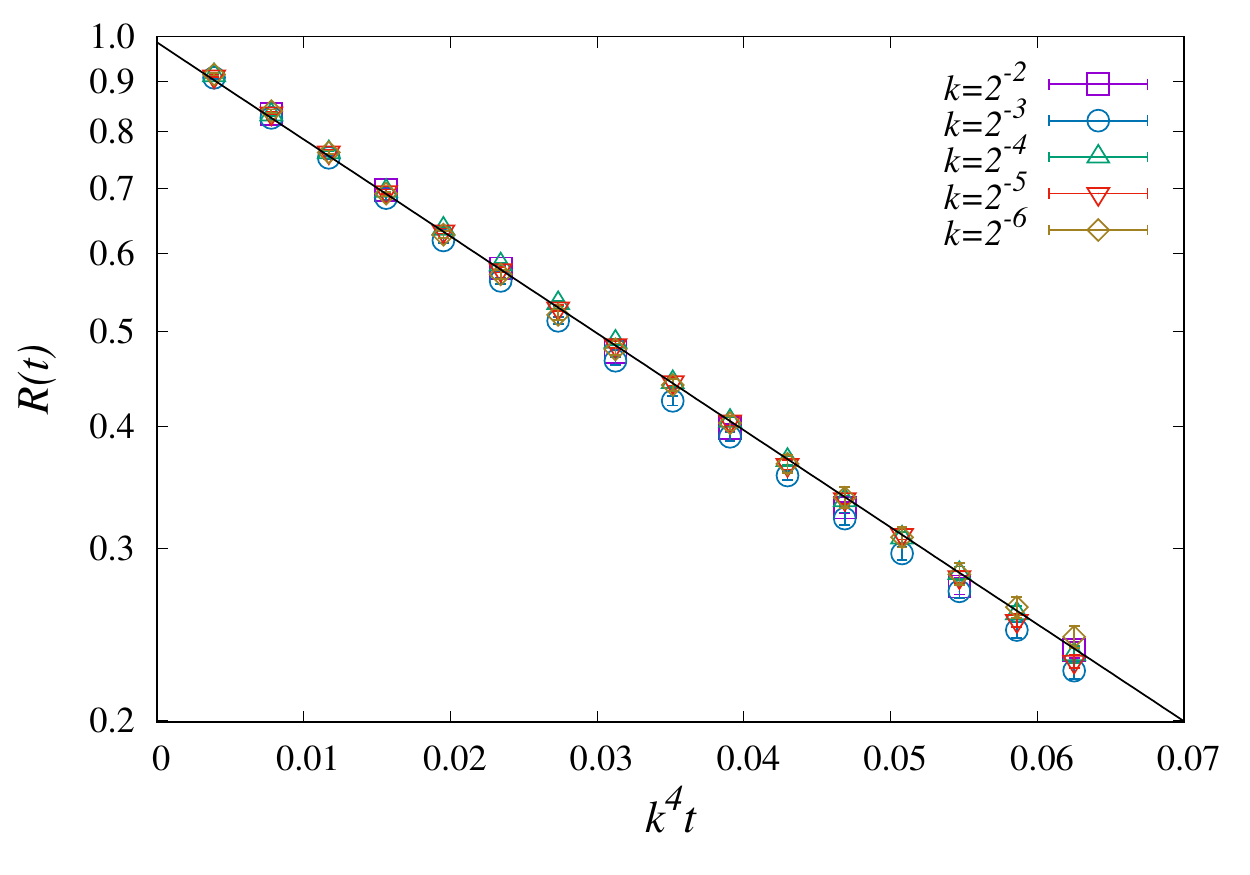}
\caption{Dynamics of the ratio $R_k(t)=n_k(t)/n_k(0)$, against $k^4t$ for various values
of $k$ and on a log-linear scale, for a $128\times128$ system of correlated walkers,
with a background base walker
density per site $n_d=3$. The solid line is a fit to a single exponential.}
\label{figsoft}
\end{center}
\end{figure}

For a function
which varies slowly as a function of $(x,y)$ (note that this implies that
the stripes in the initial condition should be wide compared to lattice
spacing), we can perform a Taylor expansion of the expression above.
We find that all terms to fourth order cancel, and the only term at fourth
order yields

\be\label{vdyn}
\pt n_{(x,y)} = -c\pxx\pyy n_{(x,y)}
\ee
where $c=1$ if following the treatment above. For further convenient reporting of the wave-vector $k$  in units of $\pi$, and taking into account in addition a factor of $4$ coming from the acceptance
probabilities in our numerical implementation, we consider instead a different normalization with $c=\pi^4/4$.
Rescaling the x-axis as in Fig.~\ref{figsoft}, this allows to recover a match to $e^{-x}$ for the fit in Fig.~\ref{figsoft}. For the rest of this manuscript, we maintain this convention for $c$.

As the stripe initial condition is a periodic square wave in $(x+y)$, it is
convenient to rewrite the above equation in the Fourier basis, and consider
only the lowest harmonic (largest wavelength)
of the square wave transform. This reduces
the dynamical equation to $\pt n_{k_x,k_y} = -ck_x^2k_y^2 n_{k_x,k_y}$.
For diagonal stripes, $k_x=k_y=k$, and $n_k$ can be exactly reduced to
$n_k(0) \exp{(-ck^4t)}$, where $n_k(0)$ is the value at $t=0$. We can now
numerically verify this behavior by calculating the Fourier ratio 
$R_k(t)=n_k(t)/n_k(0)$, and looking for a data collapse onto a single
exponential for various values of $k$. We find that $n_d=3$ provides a
sufficiently large background for a good data collapse, and show the same
for a $128\times 128$ lattice for various stripe widths (encoded in $k$)
in Fig.~\ref{figsoft} and
averaged over $80$ realizations of the stochastic dynamics for each $k$.
For smaller values of $n_d$, we find an increasing discrepancy between
different values of $k$, with the divergence growing with decreasing $n_d$.

\subsection{Discretization of continuum equation}\label{S3B}
\begin{figure}[t]
\includegraphics[width=\linewidth]{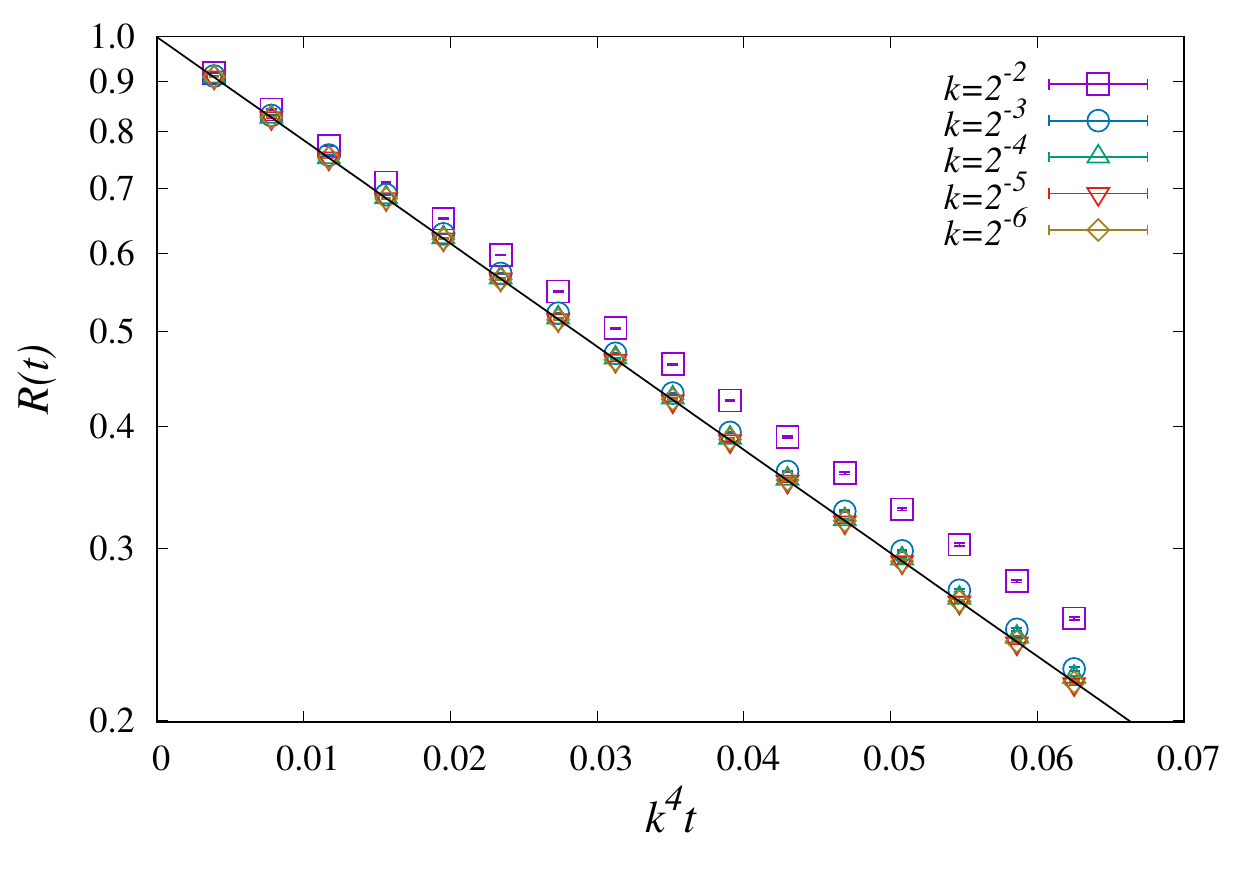}
\caption{Dynamics of $R(t)$ for a $128\times128$ system with continuum fields at each lattice site, against $k^4t$ for various values
of $k$ and on a log-linear scale. The solid line is a fit to a single exponential.}
\label{figpure}
\end{figure}

Here we study a discretization of Eq.~\eqref{vdyn} on the lattice
by defining the real valued field $n_{x,y}$, and attempt
to recover the limit of the hard-core model. From the analysis of the
correlated random walkers above, we expect that a ring exchange dynamic on
a plaquette should lead to the $\pxx\pyy$ form. To this end, we define the
``activity'' of a plaquette whose left bottom site is $(x,y)$ to be
$a=[n_{x,y+1}-n_{x,y}]-[n_{x+1,y+1}+n_{x+1,y}]$. This is evaluated
at time $t$, and the fields living on the plaquette are updated at time
$t+1$ by adding $\epsilon a$ to sites $(x,y)$ and $(x+1,y+1)$ and subtracting $\epsilon a$
from the other two sites. The constant $\epsilon$ is included to ensure that the
field remains $\in[-1,1]$, and plays only a quantitative role in the scaling
study. To deduce the dynamic rule for the activity at a single lattice site, we must consider the
four neighboring plaquettes around it which can affect the field on the
chosen site via plaquette updates. By summing $a$ for the four plaquettes
with equal weight, and carrying out a careful grouping of terms, we see that
the equation reduces exactly to the dynamical equation discussed in the
previous subsection. Using a Taylor series expansion once
again leads to a dynamical equation of the form given in Eq.~\eqref{vdyn}.
We attempt a data collapse for the decay of $R(t)$ as defined in the previous
subsection for stripe configurations using the dynamic rule mentioned above.
As shown in Fig.~\ref{figpure},
we once again find a satisfactory data collapse to
a single exponential for a large range of $k$ values.

Before we turn to the case of the hard-core model, we must note that the
dynamics described in this subsection differ in one crucial way from the
hard-core model. We can see this by considering all the plaquette configurations
which yield a non-zero value of $a$ (these are listed in Fig.~\ref{figschem}),
and observing that only the two completely ``flippable'' (or equivalently
with the largest magnitude of $a$) contribute to dynamics in the hard-core
case. This is {\it not} the case if one considers the dynamical rule used here and in the previous subsection (all plaquettes with all values of $a$ are updated). We argue in the next subsection that this leads to a strong violation of Eq.~\eqref{vdyn}.

\subsection{Continuum theory under the hard-core constraint}\label{S3C}

\begin{figure}[t]
\includegraphics[width=\linewidth]{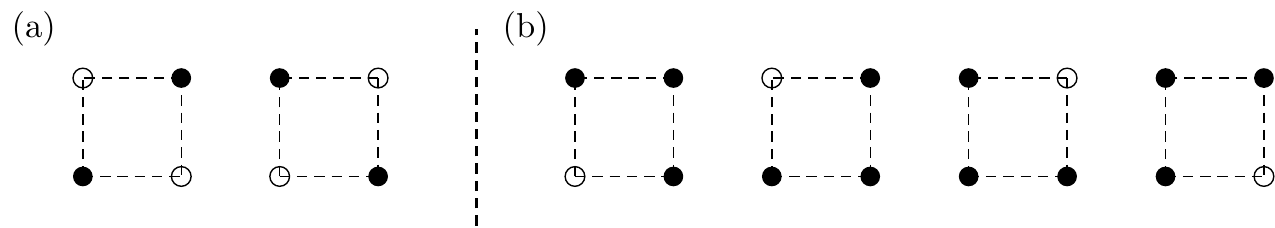}
\caption{Plaquettes which have a non-zero contribution to dynamics in Sec.~\ref{S3B}.
Plaquettes (a) are flippable, and (b) do not contribute under hard-core
discrete dynamics of Sec.~\ref{S3C}. For (b), we did not include particle-hole symmetric partners (with black and white dots interchanged) for simplicity.}
\label{figschem}
\end{figure}

Now we turn to the model described in the previous section, and restrict
particle number to at most one per site, and dynamics only to proceed via
exactly flippable plaquettes. Note that now we cannot define an updated
state using just the definition of $a$ as in the previous subsection.
We require a function which evaluates to unity only for the two flippable
plaquette configurations and zero for all others. For the purpose of this
subsection, it is more convenient to set $n_{(x,y)}=1$ for an occupied
site and $-1$ for an unoccupied site. It is not apparent if there
is a unique function which achieves this, and one of the simplest functions
which we were able to find on the plaquette to satisfy these constraints is

\begin{align}
h=\frac{1}{4}\big[&(n_{x,y+1}-n_{x,y})-(n_{x+1,y+1}-n_{x+1,y}) \nonumber \\
&-\frac{1}{4}(\Sigma_p n)
(n_{x,y}n_{x+1,y}-n_{x+1,y+1}n_{x,y+1}) \nonumber \\
&\ \ \ \ \ \ \ \times (n_{x,y}n_{x,y+1}-n_{x+1,y}n_{x+1,y+1})\big],
\end{align}

where $\Sigma_p n = n_{x,y}+n_{x,y+1}+n_{x+1,y+1}+n_{x+1,y}$ is the
sum of all $n$ belonging to the plaquette. A careful
consideration of the expression above reveals that it generates a value of
$\pm1$ for the flippable plaquettes shown in Fig.~\ref{figschem}a,
while returning zero for all other configurations
(including those in Fig.~\ref{figschem}b),
thus satisfying our requirements for a hard-core ring exchange.
The second term in the above expression is formed by noticing that the only configurations which violated
the assignment of values we desired had a difference in the types of
pair arrangements on opposite edges. Before performing approximations on this expression to derive a continuum theory, we find that it is convenient to expand the
product of the last two bracketed terms discussed above
as $(2n_{x+1,y}n_{x,y+1}-2n_{x,y}n_{x+1,y+1})$,
where we have used $n_{x,y}^2=1$ for all $(x,y)$.

For the evolution $\Delta_t  n_{x,y}$ of density of a single site, we must once again consider the
four plaquettes in which it participates. The resulting expression has a
term identical to the dynamic equation in the previous subsections,
but has an important addition in the form of the sum of
$\frac{1}{2}(\Sigma_p n)(n_{\prmx,y}n_{x,\prmy}-n_{x,y}n_{\prmx,\prmy})$
over the four plaquettes.

To get the continuum limit as done previously, we assume initial conditions with several flippable plaquettes and a smooth density profile which varies slowly spatially.
An average over stochastic dynamics and initial conditions thus allows us to
replace the terms linear in $n$ with the differential form $-\pxx\pyy g$,
where $g$ is a real valued field living in continuous space-time aiming at replacing $n$ (we
explicitly distinguish it from $n$ for this subsection due to the
correlations possibly generated by the products of $n$).

To understand the behavior of the averaged value of terms such as
$n_{x+1,y} n_{x,y+1}$, an important assumption about the correlations
has to be made. In the following of this subsection, we work in a mean-field picture ignoring correlation effects and assume that such terms can
be rewritten as the product of $g$ at the two points. In doing so,
we will be able to derive the complete dynamics only in terms of the field $g$.
This mean-field like 
assumption is true for the initial condition we work with, as it
is drawn from an uncorrelated ensemble, but it is not a priori evident if
the dynamics maintains this uncorrelated nature or rapidly builds up
correlations. The non-linear term $f$ is expressed as
\begin{align*}
&(\sum_{x,y} n)(n_{x+1,y}n_{x,y+1}-n_{x,y}n_{x+1,y+1})\\
+&(\sum_{x-1,y-1} n)(n_{x,y-1}n_{x-1,y}-n_{x-1,y-1}n_{x,y})\\
-&(\sum_{x-1,y} n)(n_{x,y}n_{x-1,y+1}-n_{x-1,y}n_{x,y+1})\\
-&(\sum_{x,y-1} n)(n_{x+1,y-1}n_{x,y}-n_{x,y-1}n_{x+1,y}),\\
\end{align*}

where the subscript below the sum indicates the plaquette over which the
sum is performed indexed by its left bottom site. Once again we expand the
products and ensure that we replace all occurrences of $n_{x,y}^2$ with unity
for all $(x,y)$. The equation above is thus reduced to a linear combination of
single body and three body terms. This can be further reduced by performing
the mean field decoupling $\braket{n_{x,y}n_{\xp,\yp}n_{\xpp,\ypp}}
=\braket{n_{x,y}}\braket{n_{\xp,\yp}}\braket{n_{\xpp,\ypp}}
=g_{x,y}g_{\xp,\yp}g_{\xpp,\ypp}$, followed by a
Taylor expansion around the relevant site
in the derivatives of $g$. As we began our analysis by considering a sum over
the four plaquettes surrounding the site $(x,y)$, symmetry restrictions
apply to the terms generated by the Taylor series, which require
that only terms with non-zero coefficients have an even number of derivatives in $x$ and $y$, and are symmetric with respect to the $x\to y$ transformation. Using this constraint and after some algebra, we find that the only surviving terms arise at fourth
order in the derivatives and that the complete dynamical equation reduces to the following expression (ignoring a global factor of $1/4$), given by 
\begin{align}\label{cfdyn}
\partial_t g = -\Big(\frac{1}{2}-g^2\Big)&\pxx\pyy g-\pyy g(\px g)^2 -\pxx g(\py g)^2\nonumber\\
&-2(\px g)(\py g)(\px\py g)-g\pxx g\pyy g.
\end{align}

The presence of non-linearities, derived even under a crude mean-field approximation, suggests that at leading order in the dynamics, the hard core constraint indeed plays an important role, and may invalidate the expectation that the coarse grained dynamics are equivalent to those of vanilla sub-diffusion.

\section{Long time persistence and eventual melt of approximate
stripe configurations}\label{S4}

We have seen in the Sec.~\ref{S2} that configurations which can be viewed
as small perturbations around a perfect stripe configuration may maintain
the memory of the initial state indefinitely. As these configurations are
highly specific, it would seem unrealistic to choose one of these as the
initial state for the dynamics of large systems. This motivated us to
study ``approximate'' stripe patterns, which are chosen to be boson
density waves with a wave-vector $\vec{k}=(k_x,k_y)$. 

\begin{figure}[t]
\includegraphics[width=\linewidth]{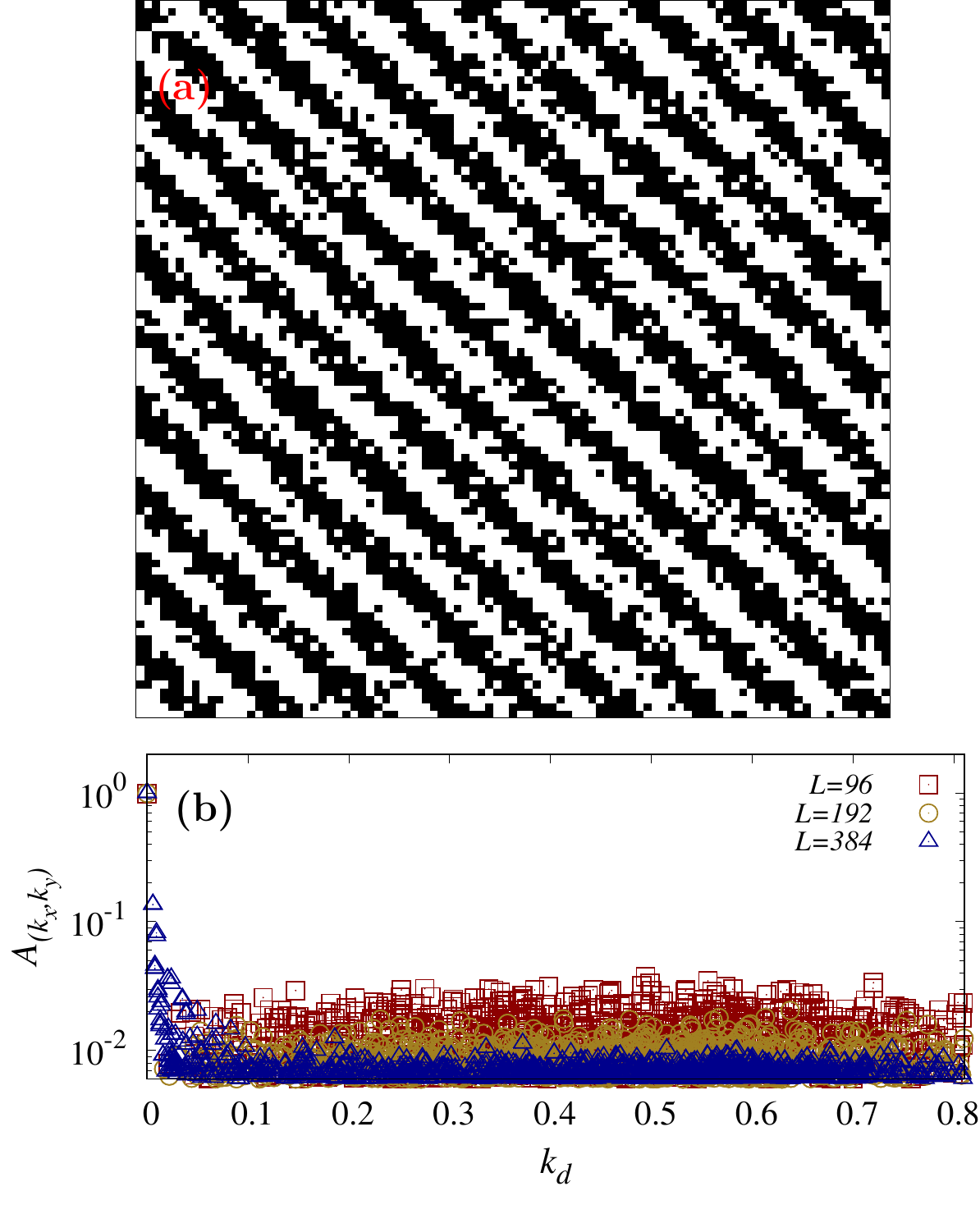}
\caption{(a) A sample boson configuration created by annealing to the target function $f_{\vec{k}}$ Eq.~\ref{tgp}
with $k_x^t=k_y^t=0.1875$ and $A=1.0$ for a $96\times 96$ 
lattice, along with (b)
Fourier components of random initial states
for larger sizes generated using the
annealing method for $k_x^t=k_y^t=0.1875$, plotted against the
nearest distance $k_d$, shows a clear peak for $k_d=0$.}
\label{figbench}
\end{figure}

\subsection{Initial state preparation}\label{S4A}

To prepare such configurations, we first
generate a target distribution using the function
\be\label{tgp}
f_{\vec{k^{t}}}(\vec{r})=A\sin(\vec{k^{t}}\cdot\vec{r}),
\ee
where $A$ takes continuous values between $0$ and $1$. We cannot generate a
configuration which has exactly this pattern as a boson configuration
can take only $\pm1$ (filled/empty) at each site.
Thus we perform a Monte Carlo simulated annealing
starting from a charge-density wave state with
$\vec{k^t}=k^t\hat{e}_x+k^t\hat{e}_y$
using an energy defined as
$E=\sum_{l}(D_l-Lf_{\vec{k^t}}(\vec{r}))^2$, where the sum is over all diagonals
and $D_l=\sum_{(i,j)\in l}\sigma_{i,j}$, is the occupancy
in diagonal number $l$. This favors an exact
match between the current and target configurations. By tuning the inverse
temperature for the annealing from $\beta=0.01$ to $\beta=20.48$ in a
geometric progression consisting of multiplying by $2$ every $10L^2$
steps, we achieve a stripe configuration which has a single Fourier
component at the target $\vec{k^t}$. We ensure that the proposed configuration
changes are long-ranged and respect the conservation laws discussed in the
previous section, thus staying in the sector where every column and every row
has exactly $L/2$ bosons. An example of a configuration created
by this procedure is shown in Fig.~\ref{figbench}a.
To check the effectiveness of the annealing procedure, we record the final
energy and find it to be $O(L)$, implying that each $D_l$ is within an $O(1)$
value of its target value, which is expected as $D_l$ is an integer and
the target value is in general a real number, and not necessarily close to
an integer value.
We find that this procedure generates a sharp peak for the desired $\vec{k^t}$
amid a weak background which decays with increasing system size. This
can be seen in Fig.~\ref{figbench}b, where we plot the Fourier component
as a function of the distance from the target peak location in Fourier space
$k_d^2=(k_x-k^{t})^2+(k_y-k^{t})^2$ for various sizes.
Due to the periodic boundary conditions in Fourier space, given by
$(k_x,k_y)\leftrightarrow(1-k_x,1-k_y)$, we consider the shortest distance to
the expected peak over the naive distance for open boundary conditions.
Note that such a configuration still contains a large number of flippable
plaquettes due to the smooth nature of the target pattern.

\begin{figure}[t]
\includegraphics[width=\linewidth]{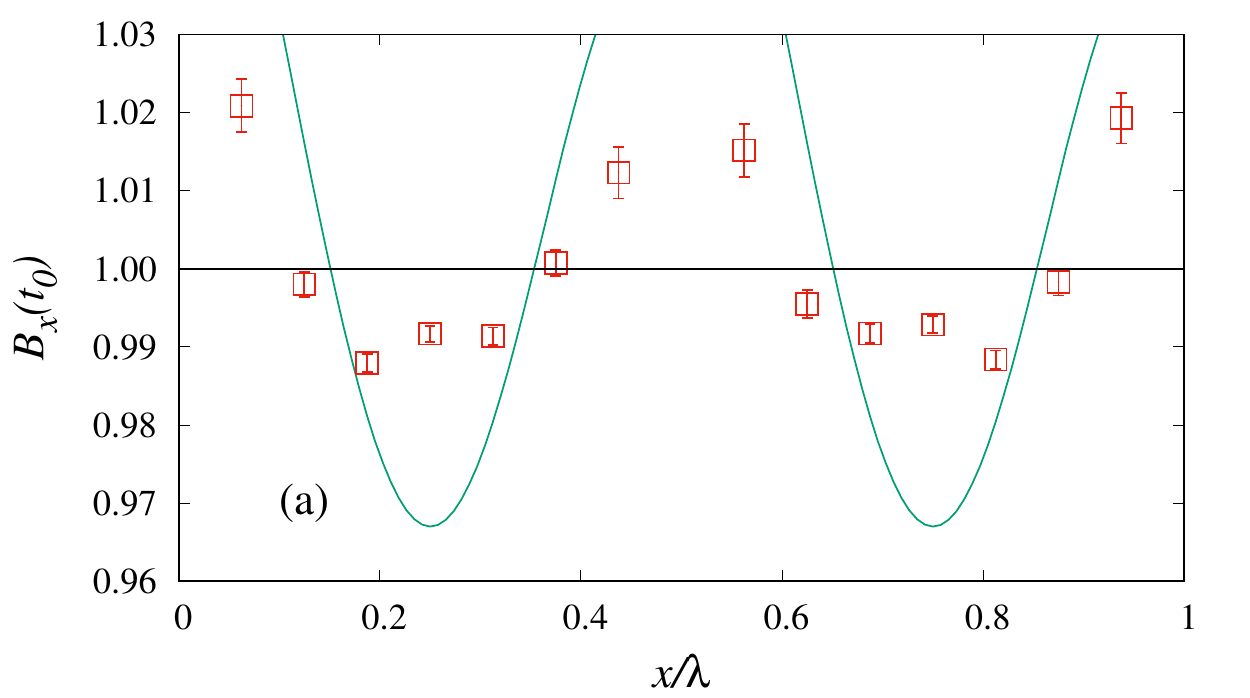}
\includegraphics[width=\linewidth]{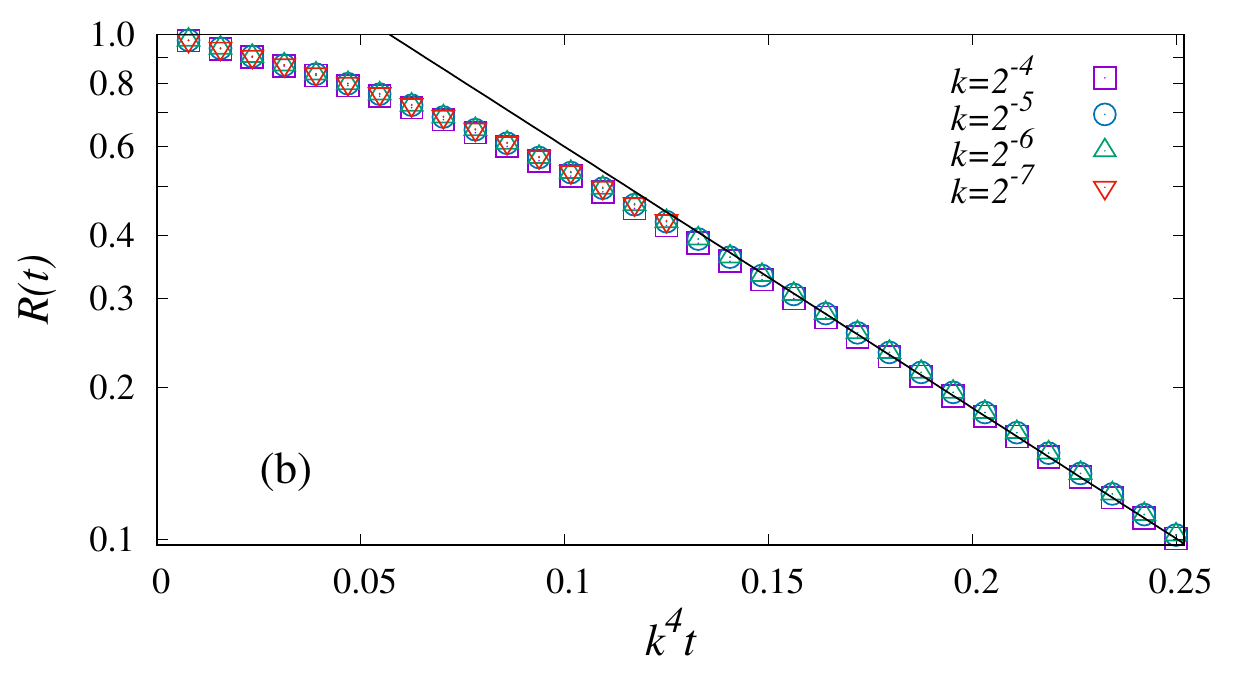}
\caption{Plots for $A=0.7$ :
(a) Comparison of average scaled profile $B_x(t)$ from microscopic hard-core
dynamics compared to continuum expression (Eq.~\eqref{cfdyn}) for wavelength $\lambda=16$
for $t_0=16$.
(b) $R(t)=n_k(t)/n_k(0)$ against $k^4t$ on log-linear scale, with long times fit to
a single exponential.}
\label{figgcomp}
\end{figure}

\subsection{Comparing with continuum field theory}\label{S4B}

In the context of the continuum field theory discussed in Sec.\ref{S3}C,
the field $g$ at time zero would now simply be equal to
$A\sin(\vec{k}\cdot\vec{r})$ with $\vec{k}=\vec{k^t}$. It is essential to numerically verify the
validity of the assumption used in developing our continuum theory,
namely that spatial correlations do not play a role in the initial dynamics. 

Using this as the initial source field in
Eq.~\eqref{cfdyn}, we find that $\pt g$ reduces to
\be
Ak^4\sin(\vec{k}\cdot\vec{r})\Big[-\frac{1}{2}+4A^2\cos^2(\vec{k}\cdot\vec{r})
-\frac{A^2}{2}\sin^2(\vec{k}\cdot\vec{r})\Big].
\ee
This immediately indicates that we have lost linearity (which would imply
no dependence on amplitude other than a global proportionality), and 
that the theory is no longer exactly separable in Fourier space.
Note here that the terms proportional to $A^2$ within the brackets oppose the
decay generated by the simple sub-diffusion, leading to a possibility of
further slowing down the dynamics. The above equation also suggests that for
$A\ll 1$, we should expect to recover vanilla sub-diffusive dynamics. For
$\vec{k}\cdot\vec{r}\in(0,\pi)$, the sign is controlled by the expression
within the brackets, and we can see that a growth of the
function can be achieved for
$\sin(\vec{k}\cdot\vec{r})<((8A^2-1)/(9A^2))^{1/2}$, provided $8A^2>1$ (that is large enough amplitude).
This condition is satisfied for $\vec{k}\cdot\vec{r}$ in the vicinity of
$0$ and $\pi$. This effect is rather non-intuitive as it implies that
the local density tends away (towards $\pm 1$) from its equilibrium
value (0) under non-equilibrium dynamics for a tunable range of
$\vec{r}$, leading to
a local reduction of entropy as the number of states available locally
reduces if we require their average to be closer to the most extreme
values which it can take.
To ascertain the extent to which our
microscopic dynamics is consistent with the continuum field theory
developed under the mean field assumption, we can now compare the
averaged value of $n_{x,y}$ against a numerical evolution of
Eq.~\eqref{cfdyn}. Looking specifically for the feature described above, we plot $B_x(t)=\langle n_{x,x}(t) \rangle / \langle n_{x,x}(0) \rangle$ in
Fig.~\ref{figgcomp}a
as a function of $x$ for the minimal period. Our initial condition 
already sets $n_{x,x}(0)=A\sin(2kx)$. We see a qualitative match to the
prediction from the continuum equation (in the sense of the sign of the difference to unity of the ratio is well captured), but a quantitative disagreement (in the amplitude of this difference)
potentially due to the build up of correlation effects beyond mean-field.

Before we move to the study of the microscopic dynamics, it is worth
considering the time scales over which the initial pattern melts under the
continuum dynamics, as done in Sec.~\ref{S2}. We show in Fig.~\ref{figgcomp}b
that the dynamics is still consistent with a single scaling variable,
given by $k^4t$, although the relaxation deviates from the single exponential
seen in the simpler cases considered in Sec.~\ref{S2}. This implies that
the additional non-linear terms in the continuum dynamics do not alter
the scaling of space-time, and this can be intuitively understood by
observing that all non-linear terms have the same order in derivatives.

\begin{figure}[t]
\includegraphics[width=\linewidth]{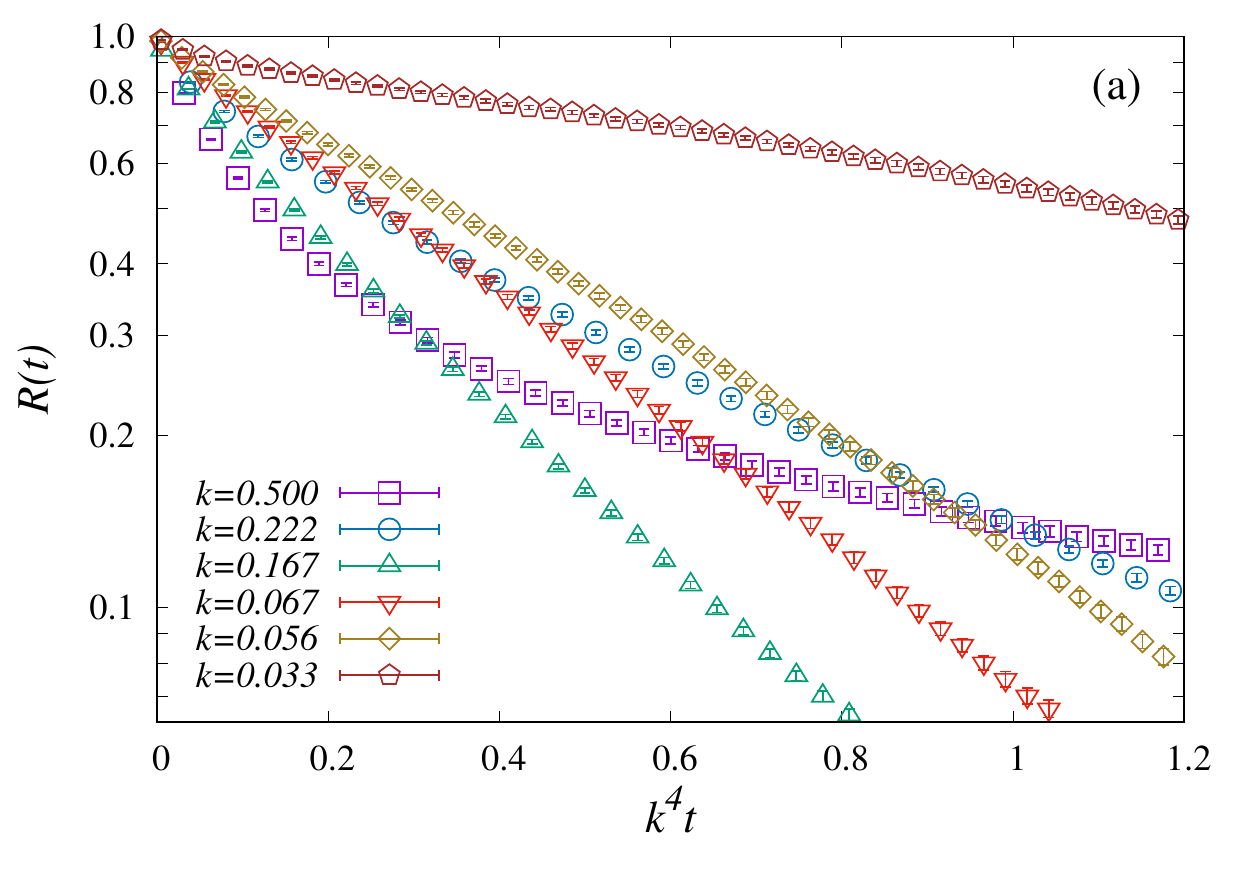}
\includegraphics[width=\linewidth]{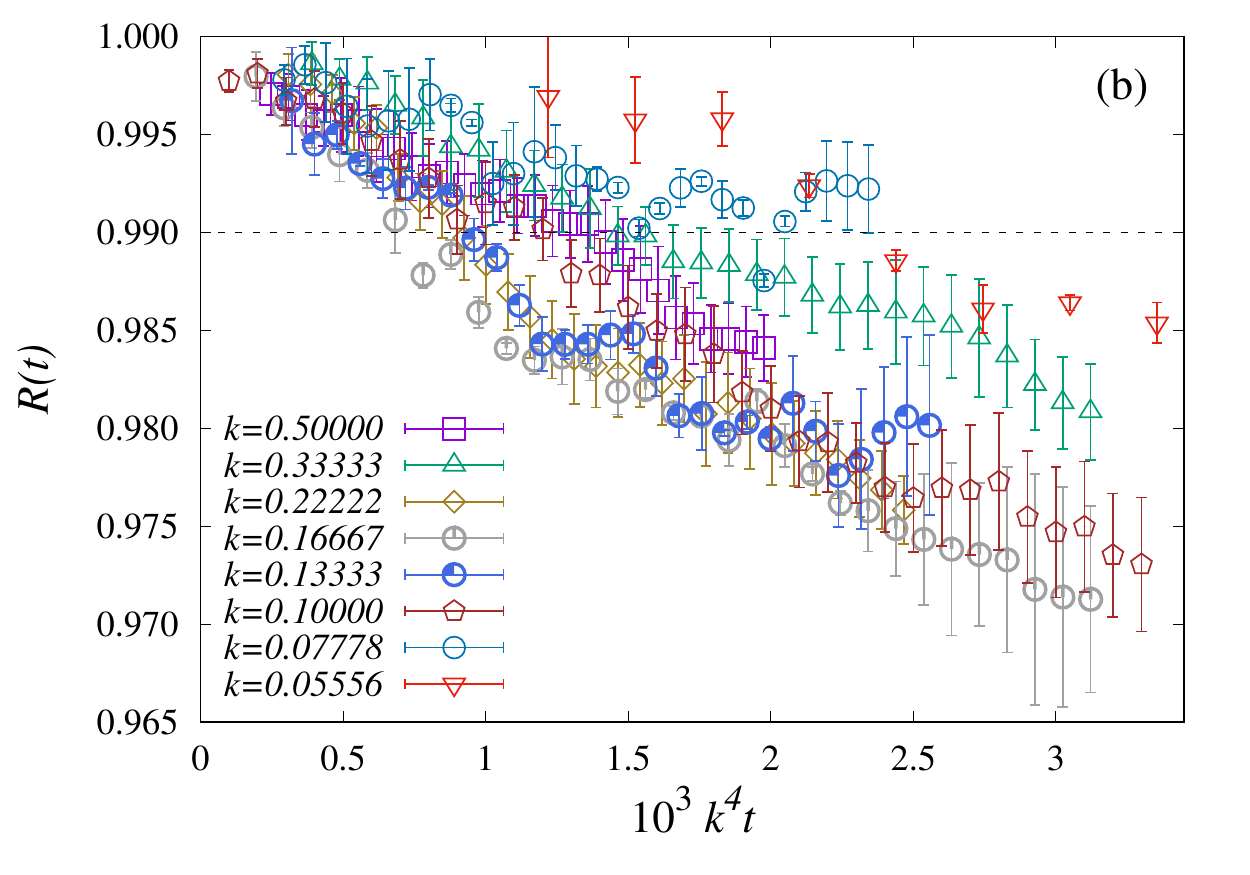}
\caption{For ``weak'' amplitude $A=0.2$. (a) Dynamics of $R(t)$ as a function of $k^4t$, in a log-linear scale. (b) Zoom on initial dynamics, that shows a consistent scaling with $k^4t$
for sample initial conditions. We do not present an average over samples as it leads 
to large fluctuations over the time axis.}
\label{figA0p1}
\end{figure}

\begin{figure}[t]
\includegraphics[width=\linewidth]{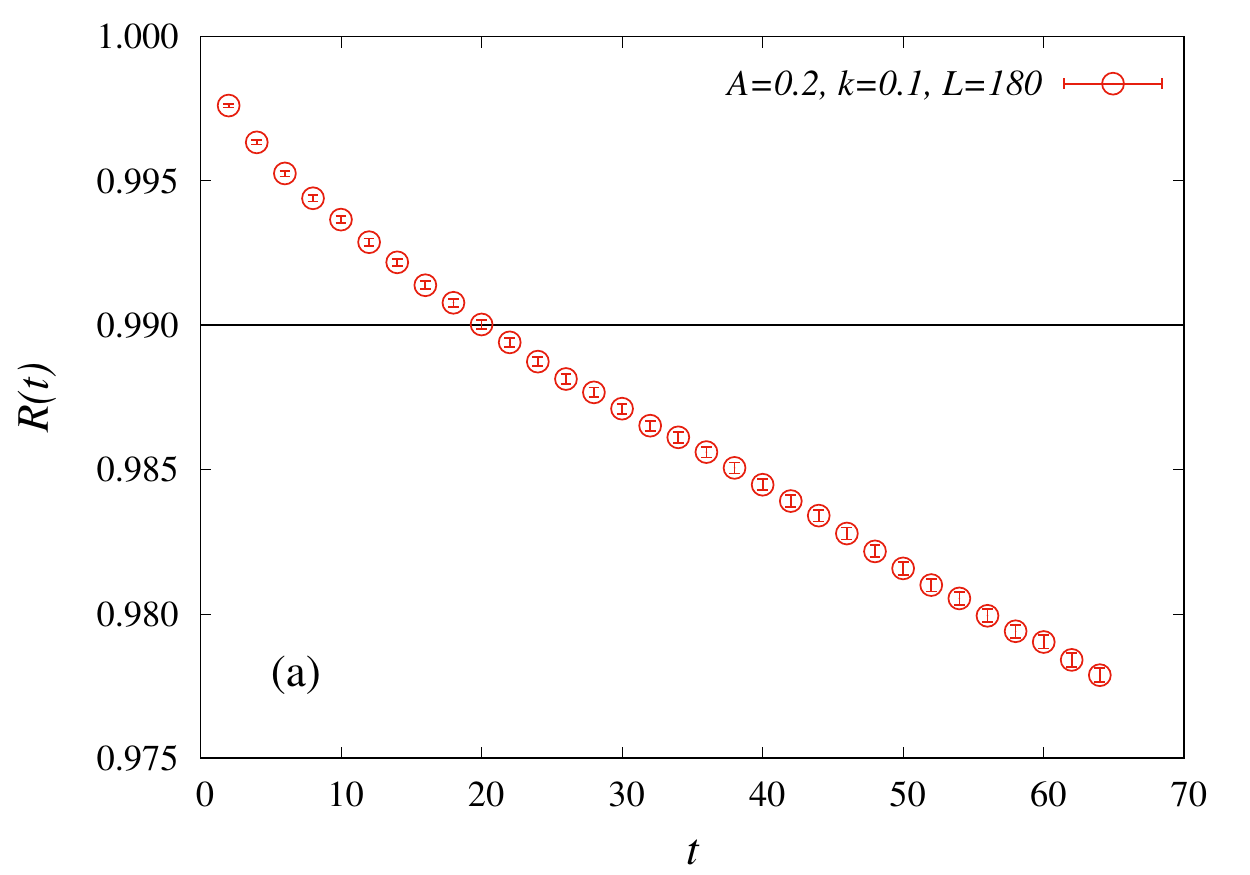}
\includegraphics[width=\linewidth]{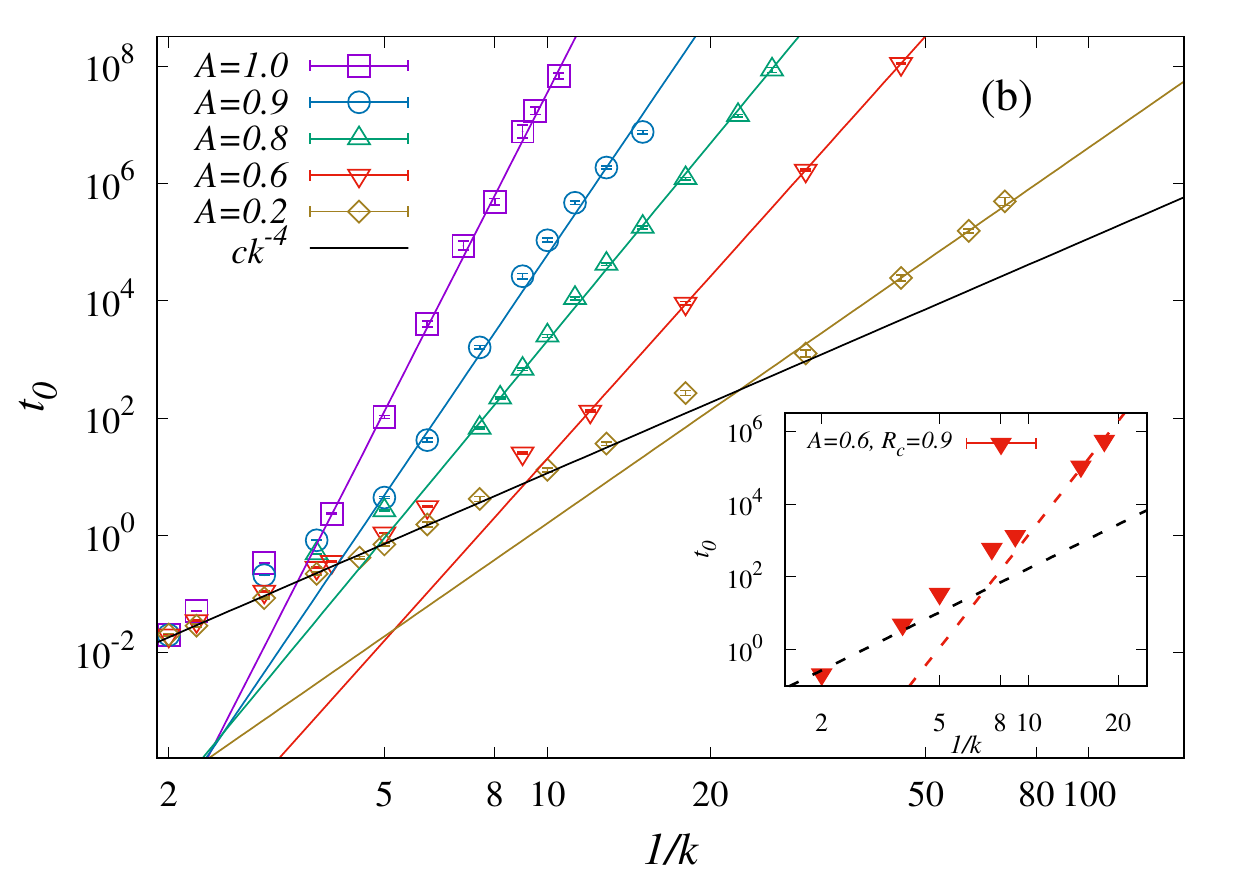}
\caption{For ``large'' values of amplitude $A$ : (a) Slow initial behavior of $R(t)$ as a function of time, for $A=0.5$, $k=0.125$ and a $L=256$ sample. To detect the beginning of the melting process, we define a threshold of 0.99 (shown by dotted line), and define $t_0$ as the intercept.
(b) Average melting starting time $t_0$ as a function of $1/k$, for
different values of $A$. The log-log scale emphasizes the power-law dependence. The black solid line corresponds to a $1/k^4$ power-law.}
\label{figtexp}
\end{figure}

\subsection{Exact numerical evolution:
Beginning of melting}\label{S4C}

We now study the dynamical behavior starting from a configuration generated
by the method discussed above. To carry out an analysis of the evolution
of the coarse-grained structure of the initial configuration, we choose to
study the decay of the dominant Fourier component via the already
defined ratio $R(t)=\frac{n_{\vec{k}}(t)}{n_{\vec{k}}(0)}$. 
To get an intuition about the
effect of lattice spacing, we first consider the decay of $R(t)$ over
a range of decreasing values of $k$, starting from $k=1/2$. As shown in
Fig.~\ref{figA0p1}a, for $A=0.2$, there are clear deviations away from a single
exponential for large values of $k$, whereas intermediate values of $k$
appear to agree partially, and for small values of $k$, we once again deviate from the
expected $k^{-4}$ scaling. To study the initial stages of the decay of the
wave pattern, we can look at $R(t)$ in the regime where it is close to
unity. Here we find a collapse consistent with a time scaling with $k^{-4}$ for a
large range of $k$, including large $k$ values (as shown in Fig.~\ref{figA0p1}b). This is quite surprising, because the regime of short-time scales and large-wave vectors is not the one where we expect the hydrodynamic prediction of Sec.~\ref{S3} to hold. We have no simple explanation for this observation.

For larger values of $A$, we observe a completely different behavior. We find that $R(t)\approx 1$
for a non-trivial amount of time after the initialization of the dynamics, an illustration of this behavior of $R(t)$ is given in Fig.~\ref{figtexp} for $A=0.5$. We define the beginning of the melting process by the
first time $t_0$ at which $R(t)$ crosses $0.99$. This threshold is chosen
arbitrarily and a different threshold does not change the result
qualitatively (as shown using a threshold of $0.9$ in the inset
of Fig.~\ref{figtexp}b).
We study this for a few values of $k$, and for $20$
realizations of the initial condition for each $k$. In addition to this,
for each realization, we run sufficient number of realization of the
stochastic dynamics to ensure that we get a good estimate for $t_0$.
We have taken systems of linear size in the range of $150-250$,
as the property of self-averaging allows us to narrow the spread in the values of $t_0$.
We find a strong dependence of the averaged $t_0$ on the inverse wave-vector
$1/k$. A linear fit on a log-log scale of $t_0$ vs $1/k$ reveals various
power law regimes (see Fig.~\ref{figtexp}b) that depend on the amplitude $A$, the most extreme of which is achieved for $A=1$, where
the dependence is $t_0\sim (1/k)^{18.1(3)}$. Such a strong dependence on the
initial pattern suggests that the mechanism for melting is initiated by
a coordinated rearrangement of bosons which is favorable for dynamics.
Note that for all values of $A$, there exists a window of $1/2>k>k^0(A)$,
where
we find good agreement with $(1/k)^{4}$, as predicted from the continuum theory, with $k^0(A)$ reducing with reducing $A$.

\subsection{Exact numerical evolution:
Post-melt scaling}\label{S4D}

The continuum theory developed in Sec.~\ref{S3}, which can be valid only for small wave-vectors and long times, cannot describe the results in Fig.~\ref{figtexp}b and the period for
which we observed $R(t)\approx 1$, and this suggests that it is not the
appropriate theory to understand this ``prethermal'' behavior. 

\begin{figure}[t]
\includegraphics[width=\linewidth]{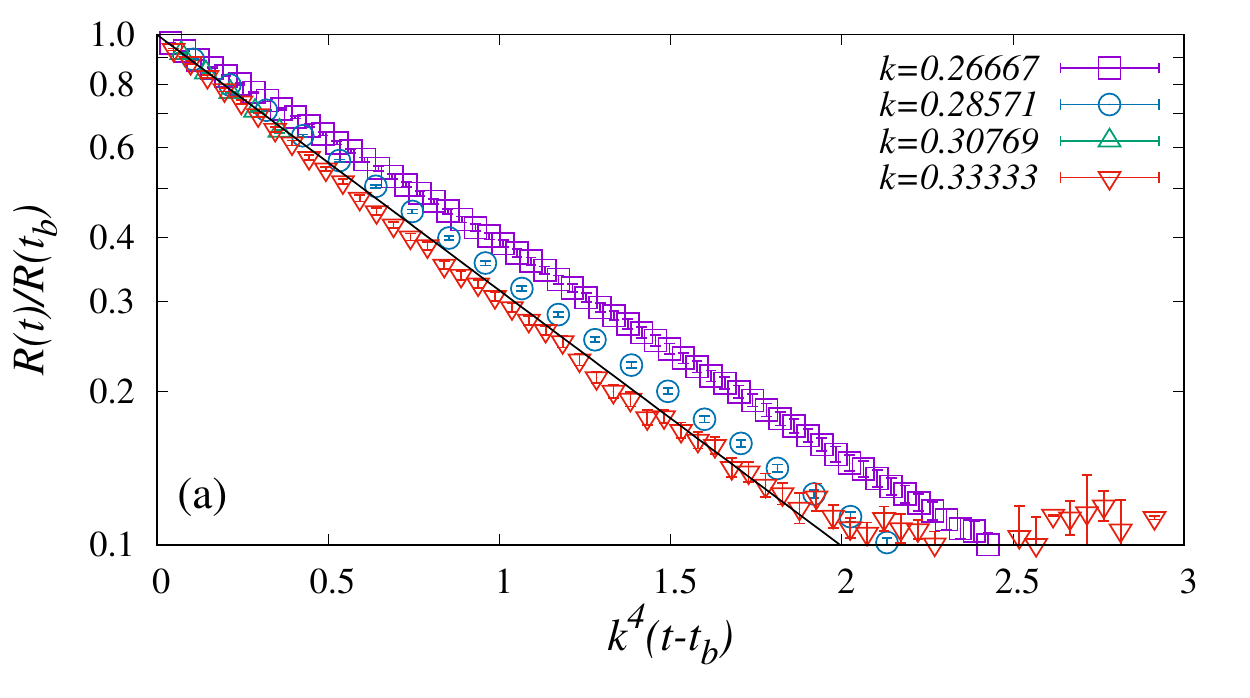}
\includegraphics[width=\linewidth]{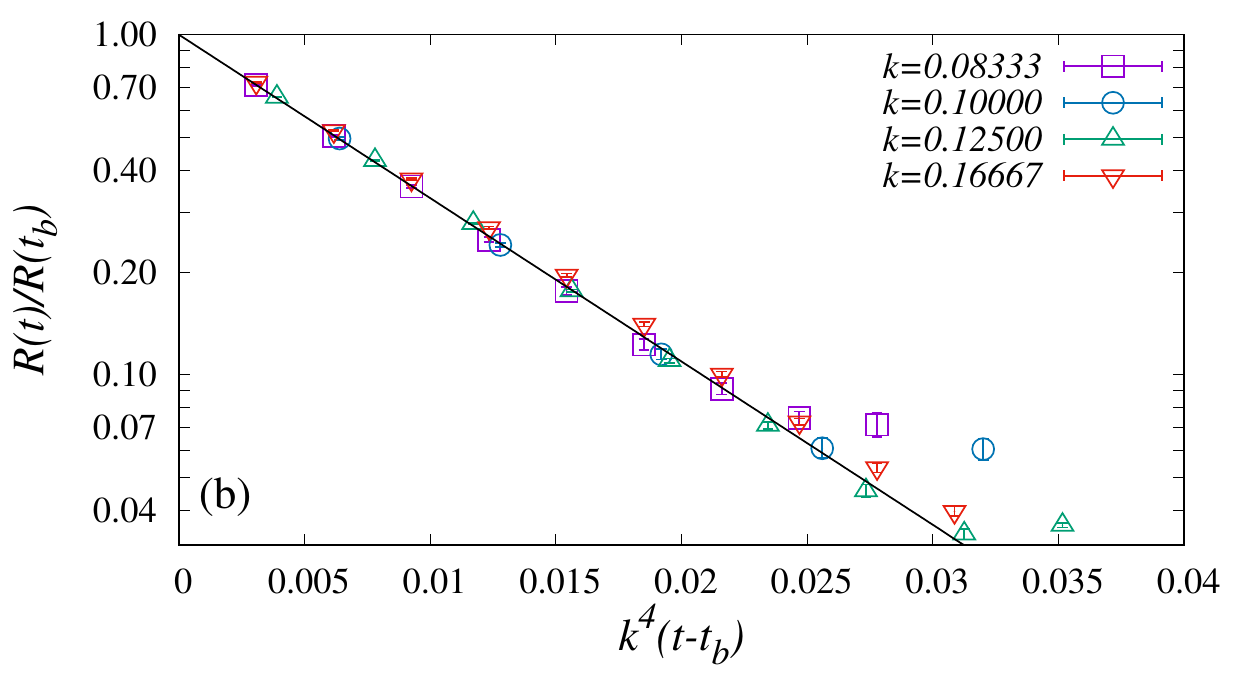}
\caption{$R(t)$ scaled by the value closest to $0.01$ ($R(t_b)$)
in our dataset vs $k^4 t'$ where $t'=t-t_b$ is the time shifted by the reference time $t_b$,for different values of $k$. A log-linear scale is chosen to emphasize the expected scaling in $k$ for different values of $k$. (a) : For the main ring-exchange model discussed in the manuscript. (b) Similar scaling for an extended model where we allow ring-exchanges models around $2\times 1$ and $1\times 2$ plaquettes.}
\label{figmelt}
\end{figure}

We can check however if the continuum prediction is upheld after the melting process
with a simple scaling collapse. We first define a time $\tp=0$, which is
taken to be the well after the beginning of the melting process, as the time at which
$R(t)$ drops below a threshold of $0.01$ (chosen just for the convenience
of analysis, once again we checked that this value only plays a quantitative
role). We can now attempt a parameter free scaling collapse of $R(\tp)$
and would expect to get a linear profile on a log-linear plot for a
range of $k$ by scaling the $\tp\to k^4\tp$. We present this analysis
in Fig.~\ref{figmelt} for $A=1.0$
and find a reasonable agreement with the expectation developed
above, even if not entirely satisfactory. To further check the applicability of the continuum theory, we also
consider ring exchange dynamics over $2\times 1$ plaquettes, which we expect to the same Eq.~\eqref{vdyn}, but have much short pre-thermal times (as naturally be expected from a longer-range type of exchange). This allows to probe a larger range of wave-vectors $k$ values, as presented in Fig.~\ref{figmelt} where the adherence to a $k^4 \tp$ is clearly improved.

\section{Pattern of dynamical activity in prethermal and melting
process}\label{S5}

\begin{figure}[t]
\includegraphics[width=\linewidth]{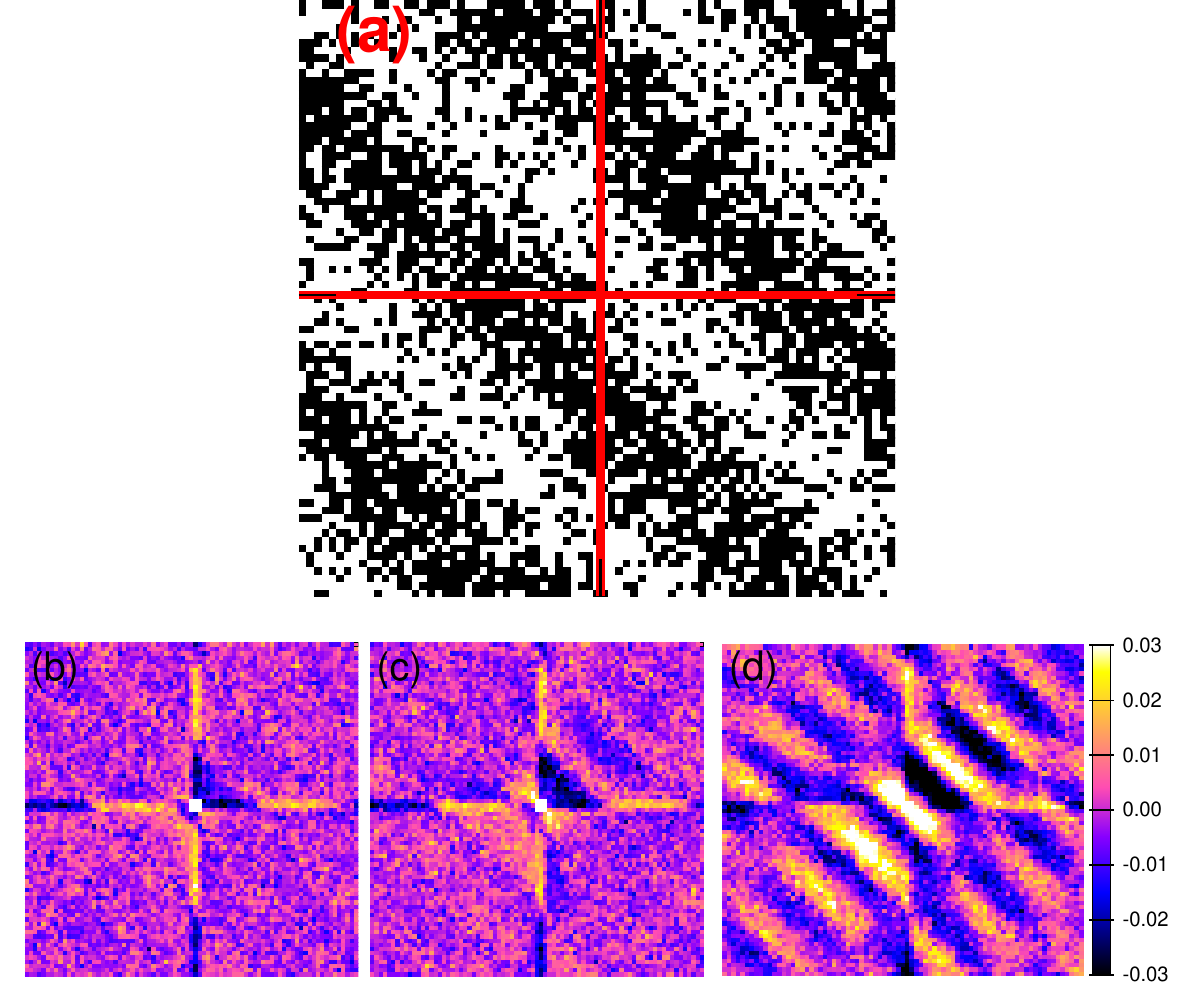}
\caption{(a) A sample initial configuration for an $80\times 80$
lattice with $k=0.05$ and $A=0.6$. Intersection of axes indicates origin.
(b-d) Flippability correlator $C(x,y,t)$ for the same system size, averaged over $80$
initial conditions and $128$ realizations of the dynamics for each,
at times (left to right) $t=2^{17},2^{19},2^{21}$.}
\label{figCwave}
\end{figure}

The previous section shows that a mean field treatment is unable to
recreate the slow dynamics seen numerically, thus hinting at the presence
of correlations which are neglected at a mean field level.
To understand the large-scale mechanisms involved in the prolongation of
the prethermal lifetime and the onset of melting, we consider the build
up and correlations of flippable plaquettes. This serves as a proxy for
identifying dynamically active regions and their evolution. We study real-space
 flippability correlations through the normalized connected correlator
$$C(x,y,t)=\frac{\braket{P(0,0,t)P(x,y,t)}-\braket{P(0,0,t)}
\braket{P(x,y,t)}}{\braket{P(t)}^2},$$ where $P(x,y,t)=1$
if the plaquette whose left bottom site is $(x,y)$
is flippable and zero otherwise, and $P(t)$ is the spatially averaged density of
flippable plaquettes. Large (low) values of $C(x,y)$ at a given time will thus indicate strong (weak) correlations of flippability with the initial point. Recall that we have chosen starting
configurations from annealing to a potential which is only a function of
the tilted coordinate $x+y$. This means that we have a freedom of choosing
the origin for our correlation function at any $x-y$ for fixed $x+y$.
We take advantage of this symmetry by averaging over all equivalent positions
of the origin. For the $x+y$ position of the origin we use our potential
function as defined in Eq.~\eqref{tgp} and set our origin to be the point
satisfying $\vec{k}\cdot\vec{r}=\pi/4$. We make this choice as it lies at the
threshold between highly active and inactive regions, defined by $\braket{n}=1/2$
(medium boson density, high density of flippable plaquettes) and $\braket{n}=1$
(high boson density, low density of flippable plaquettes).

\begin{figure*}[t]
\includegraphics[width=\linewidth]{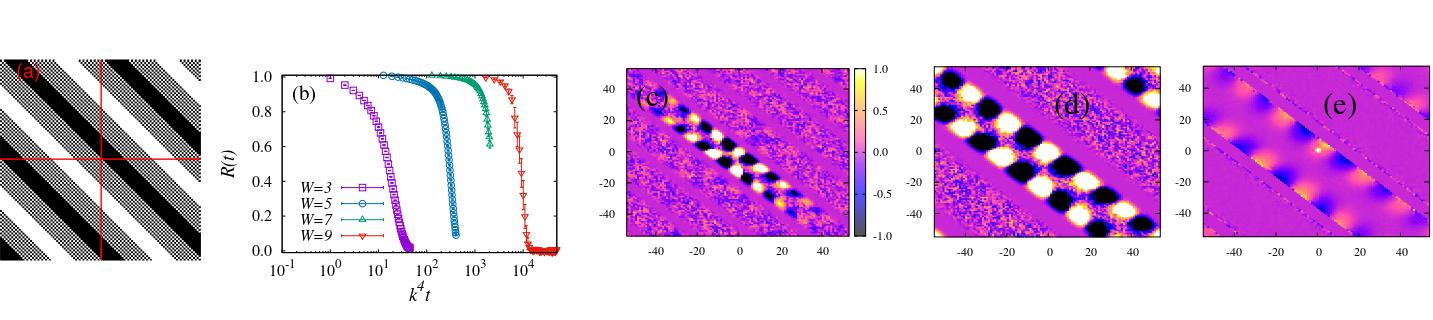}
\caption{
(a) A sample filled-staggered-unfilled-staggered type initial configuration, as described in the main text, with $W=13$ consecutive diagonals filled.
(b) $R(t)$ at the dominant wave-vector for various values of $W$,
against rescaled time $k^4t$.
(c) Flippability correlator $C(x,y,t)$ for configuration in (1) at $t=2^{23}$
(d) $C(x,y,t)$ for $W=27$ at $t=2^{21}$ with the same conditions.
(e) Same as plot for $W=27$, scaled instead by $\langle P_{d_i}\rangle\langle P_{d_j}\rangle$.
}
\label{figsqne}
\end{figure*}

\subsection{Density wave patterns}\label{S5A}

We begin by considering the initial conditions studied in the previous sections.
To ensure that we are able to observe large scale (slowly varying spatial)
features in the correlation function, we choose a system size of
$80\times 80$ and $k=0.05$ at $A=0.6$
(a particular realization of this is shown in
Fig.~\ref{figCwave}). To gather high quality statistics we
average over 80 realizations for the initial condition and for
each realization we run 128 realizations of the stochastic dynamics. The time $t_0$, which denotes the
beginning of the melting process, is $\approx 10^5$ or $2^{17}$ for $k=0.05$,
as seen in Fig.~\ref{figtexp}. We study $C(x,y,t)$ up to time scales
of $2^{21}$ and find that strong anti-correlations develop at short range
for the density of flippable plaquettes at early times, and sustain until the
intermediate stages of melting. This is shown in
Fig.~\ref{figCwave}, where snapshots of $C(x,y,t)$ at
$t=2^{17}\to2^{21}$ are presented. 

This profile shows the development of active regions
(bright) surrounded by inactive regions (dark), and suggests that
a mechanism of excluded dynamical regions may be the source of dynamical
slowing down. Note also that Fig.~\ref{figCwave} suggests a bias of the
dynamics towards the $x$ and $y$ axes, which may be expected from the
continuum theory as well, due to the lack of radial symmetry. We see that
even before the melting time, weak correlations are
already built up along the $x$ and $y$ axes, but in a manner which is
anti-correlated and modulated with the approximate stripe width. We find that
this pattern begins to appear at times as short as $t=2^4=16$, but we present
data at the last possible time value before melting
which we have recorded as the pattern is
visible with significantly more clarity. 

Another important aspect of the correlation
pattern to note here is the dependence on the $x-y$ coordinate (``perpendicular'' to the initial wave-pattern), which is not
built in in the initial condition as the energy used in the annealing
process used to generate the configuration has no $x-y$ dependence. This behavior cannot also be expected from the continuum theory since we start of with a single wave like configuration with $k_x=k_y$. 

We find that an important condition for the existence of such correlation patterns is that they are specific values of $A$ where the dynamics is not described by simple sub-diffusion. For example, we studied the case of $k=0.05$ at $A=0.2$, which shows conventional dynamics (as far as we can see from Fig.~\ref{figtexp}), and did not find any non-trivial correlations down to a precision of $10^{-4}$ in $C(x,y,t)$. This is as expected from a mean-field treatment, as it forbids correlations by definition.

The evolution of the correlation landscape for the duration of the melting process presented in Fig.~\ref{figCwave} suggests that it proceeds through a merging of dynamically active regions. This will be discussed in more detail for step-like initial conditions, where a complete melting process can be observed with higher clarity and on longer times. 

\subsection{Maximally active square wave initial conditions}

To gain a better understanding of the melting process, we begin with a more artificial initial
condition chosen to have regions of flippable plaquettes with maximal and
minimal density. We consider
alternating patterns such as the one shown in Fig.~\ref{figsqne}, with $W$
(necessarily odd)
consecutive diagonals filled, followed by $W+1$ in a perfect staggered (``N\'eel") pattern, followed
by $W$ empty diagonals, and finally by another $W+1$ in a N\'eel pattern to
close out one period. These constraints are chosen to ensure that a filled
region is bounded by empty diagonals and vice versa.
The lattice size is then $N(2W+2(W+1))$, where $N$ is
the number of periods. Although these configurations show melting times
(quantified by the decay of the appropriate Fourier component) which are faster
than those studied above, they show a scaling fore pre-melting which is slower than the 
conventional $k^{-4}$, as seen in Fig.11. We find that the flippability correlator $C(x,y,t)$ for these
unconventional initial conditions shows an extension of the anti-correlated
pattern in the $x-y$ direction with a periodic modulation across distances
large compared to lattice spacing.
This is seen clearly for $W=13$ with $N=2$ at time of
$t=2^{23}=8\times 10^6$ in Fig.~\ref{figsqne}.

\begin{figure*}[t]
\includegraphics[width=\linewidth]{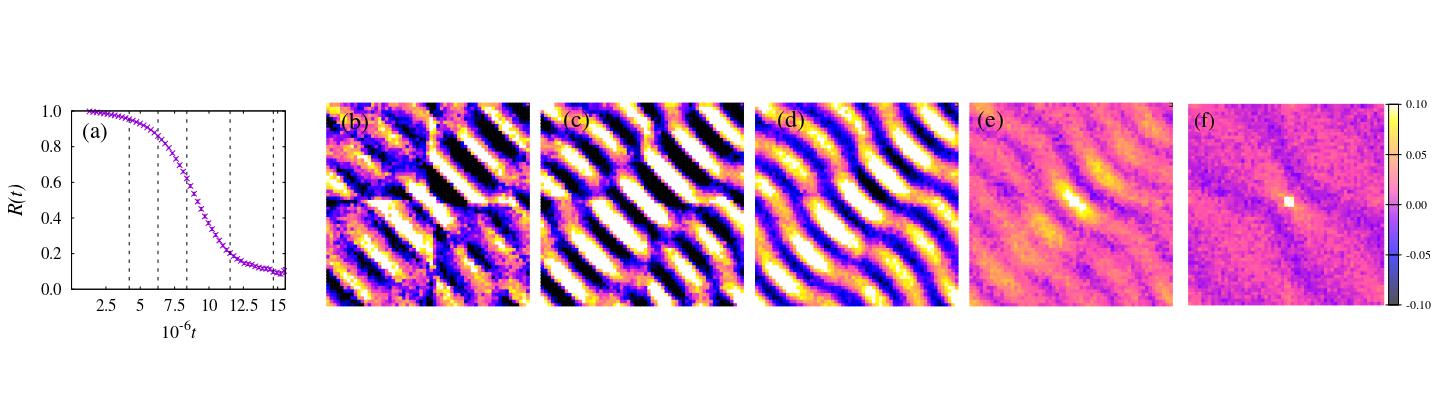}
\caption{(a) $R(t)$ for $W=7$ and 2 stripes with vertical lines
marking the times (b-f) $[4,6,8,11,14]\times 2^{20}$ for which
$C(x,y,t)$ profiles are collected.}
\label{fig7melt}
\end{figure*}

An important characteristic of the pattern described above is the value of
the wave-vector associated to this pattern. We find that this is determined by the approximate stripe
width, i.e. the wavelength $1/k$. Our data suggest that this occurs due to
a development of the correlation along the $x$ and $y$ axes, which is
limited by the size of active regions (high density of flippable plaquettes).
Growth of correlation within an active region is mediated by alternating patterns
of correlation and anti-correlation, which can be thought of as being generated
by reflections off of the boundary between active and inactive regions, as
shown in Fig.~\ref{figsqne}. Repeated reflections
create the observed periodic pattern, thus linking the periodicity and the
width of the initial stripes. This is clearly seen in the approximate
periodicities of the pattern for $W=13$ and $27$ in Fig.~\ref{figsqne}
where the profiles look similar even though the stripe width is
changed by a factor of 2.
Strong signatures of the periodic pattern
for stripe widths (or alternatively wavelengths) of $13$ and $27$ at
approximately the same time indicates the emergence of this pattern at
a time scale which only weakly depends on the wave-vector $k$,
and that it develops at a time scale parametrically much
smaller than the melting time.
Note that an extrapolation based on Fig.11b indicates the beginning of
melting to occur earliest at $t\approx 2^{27}$, assuming a lower bound of the scaling behavior as $k^{-4}$ (which is clearly slower than the scaling our numerics show in
Fig.11b).

To develop an estimate of the relevance of
this patterning to the process of melting, we first remove the normalization of
$\braket{P(t)}$ which we have absorbed into the definition and instead
look at $C(x,y,t)$ scaled by
$\langle P_{d_i}(t)\rangle\langle P_{d_j}(t)\rangle$, where $\braket{P_{d_i}(t)}$
is the average number of flippable plaquettes in diagonal number $i$ 
at time $t$.
This makes the correlation function effectively lie between $-1$ and $1$,
with either limit describing a saturation to the largest (smallest) values.
In particular, a value of $-1$ implies that the density of flippable
plaquettes is zero, leading to a complete arrest of the dynamics. This
version of $C(x,y,t)$ is plotted in Fig.~\ref{figsqne}e and we
see clearly that close to the frontier of dynamical activity, we do get
a complete anti-correlation around the reference region. This suggests that
the dynamical activity is likely strongly controlled by the unconventional
pattern in the $x-y$ direction. 

Now we turn to an investigation of the role of the periodic patterns observed
above in the melting process. We can study this by choosing a small enough
$W$ so that the melting process is captured within the time scale of $10^9$
which we can simulate. We find that for $W=7$, the ratio $R(t)$ begins to deviate from
unity at a time $t_0 \sim 2^{21}=2\times 10^6$, and reaches a value of 0.01 by
$t_b \sim 2^{24}=1.6\times 10^7$ (data plotted in Fig.~\ref{fig7melt}).
In this interval, we find at time $t=2^{20}$ a formation of
stable periodic structures in the $x-y$ direction, and see a connection of the
dynamically active regions for $t=2^{22}$. Finally by $t=2^{24}$, the
correlation pattern has relaxed into a single wave in the $x+y$ direction,
restoring the symmetry in the $x-y$ direction expected from the continuum
prediction. The role of correlations is expected to be insignificant 
past this time point.

\section{Summary and outlook}\label{S6}

We studied a system of hard-core bosons on a square lattice evolving
under classical stochastic dynamics using ring exchanges. We find
that boson density waves motivated by the patterns present in the Hilbert
space fragmentation of this model approach equilibrium on an extremely
long time scale, which diverges with the inverse wave-vector as
$k^{-\alpha}$ for small $k$,
where $\alpha$ depends on the amplitude $A$ of the initial
density wave configuration. For decreasing values of $A$, we find increasingly
larger windows in the {\it large} $k$ regime where we observe a scaling of
$k^{-4}$. Our numerical studies and mean-field treatments of simplified models
show the emergence of sub-diffusion on the other-hand in the hydrodynamic low $k$ limit. For the hard-core case, we derive a dynamical equation using a mean-field like assumption which is able to capture partially the unconventional feature of an
increase in slope around the nodes of the pattern driven by the non-linear terms
in the dynamical equation but which also predicts a time scaling of $k^{-4}$. This leads us to the conclusion that the microscopic dynamics are controlled by strong correlations which are built up in the prethermal regime, which cannot be accounted for by this mean-field treatment.

To understand the mechanism driving the long prethermal regime,
we study the correlation function of flippable plaquette density for
different instances in time. We find that a strong anti-correlation is built
up in the direction orthogonal to the modulation direction for the initial
condition, even extending to the entire length of the diagonals for
intermediate system sizes. We thus observe the formation of
isolated dynamical puddles and find that the mechanism leading to the
melting is characterized by a merger of these puddles, which relax into
a profile which is consistent with what we expect from the continuum theory.
Thus, we have found an example where correlations control crucial aspects of 
the dynamical behavior, and a simple mean-field like hydrodynamic description is
insufficient. An important question remains: how generic are the long melting processes that we observe, 
in particular to which additional classes of initial conditions
and types of dynamics do they apply?
From the discussion of perfect stripe configurations
presented in Sec.~\ref{S2}, it is easy to see that similar results
can also be expected for stripes which have a different tilt,
as long as $k_x,k_y\neq 0$. Regarding the possible set of dynamics
which can be expected to engineer a similar phenomenology, we
observe that a restricted spread of excitations similar to
fractonic systems \cite{nandkishore2019fractons} may be at play
here. For the specific model we study here this can be seen
by considering the example discussed in Sec. \ref{S2}, where
we see that the plaquette excitation is not able to move freely
in the direction where there is a lack of similar excitations.
More complex versions of this example can be generated
by other spatially-larger dynamical terms conserving dipole moment,
and can be expected to have a similar constraint when diffusing
into spatial regions which lack dynamically active cells.
This mechanism is nevertheless not as cleanly formulated and generic as the ones found in fractonic models
\cite{nandkishore2019fractons,chamon2005quantum,pretko2018fracton}. In particular it is still possible to define
special particle configurations in our model, in which an active
plaquette can move locally without restriction.
However, our numerical results and the intuition developed in
Sec.~\ref{S2} suggest that for modulated structures which vary
over large length scales,
the phenomenology which we have discussed should hold for generic
dipole moment conserving local dynamics.

We have shown that the decay of the overlap with the initial state with time
shows a similar behavior upon comparing the exact quantum dynamics and our classical automaton for a small system size. This leads us to
speculate that the behavior studied in the automaton language
might inherit properties which can also be relevant for the exact quantum
evolution for long times, as already seen for the study of sub-diffusive
behavior in 1D \cite{RefFP}. First of all, we expect that our hydrodynamic description might be equally valid also for the quantum dynamics, in case the quantum system exhibits a hydrodynamic long-time behavior. This would imply sub-diffusion also in the quantum case. Further, it might be possible that the onset of hydrodynamic behavior also in the quantum model might be delayed to very long times. This would have the consequence that a large intermediate time window exists with unconventionally slow dynamics.
As this model is related to the quantum
link model \cite{karpov2021disorder} via a relaxation of charge
conservation\cite{RefC}, similar physics may be expected to emerge there
in a more restricted setting.
The category of initial conditions which we have studied may be
seen as high energy states of a quantum XY model with ring exchange
interactions \cite{sandvik2002striped}. We expect that an initial state with a fully staggered charge density pattern, located at low-energies for such a model with large ring-exchange, would relax extremely quickly as it has the maximal possible density of flippable plaquettes by the ring-exchange term.

Given that the stripe configurations we consider are qualitatively
similar to domain walls in some ferromagnetic systems (such as found {\it e.g.} in Ising models~\cite{spirin2001freezing}), it is important to note
that the stability of domain walls in such cases is ensured by energetic reasons and is only expected at low temperature.
In contrast, we do not have any such potential term which reinforces
local correlations, and our long lived stripe structures are generated
purely by the restricted dynamics available to the system.

The crossover
between Hilbert space fragmentation, which dictates preservation of the
memory of the initial state to arbitrarily long times, and the long
prethermal plateaus we see here, is another promising direction in which
investigations can be carried out to better our understanding of the
processes involved. Although we do not find any exact
conserved quantities which could explain the slow relaxation seen
in our dipole conserving model, another interesting follow-up would be to try to build a theoretical
framework in terms of statistically localized integrals of motion \cite{rakovszky2020statistical}.

Finally, the rapid increase of the lifetime of the
prethermal behavior with inverse wave-vector suggests that rare events,
which create configurations allowing the applicability of the hydrodynamic
theory, play a key role. An improved understanding of the potentially
large deviations which lead to the above mentioned phenomenon would help
greatly in developing a coarse-grained description of the dynamics at the
threshold between the prethermal and equilibrium regimes.

\begin{acknowledgements}

We would like to thank Cl\'ement Sire, Kedar Damle and Frank Pollmann for fruitful
discussions. Computational resources for this project were provided by
LPT Toulouse and MPIPKS.
This project has received funding from the European Research Council (ERC) under the European Unions Horizon 2020 research and innovation program (grant agreement No. 853443).

\end{acknowledgements}

\appendix

\section{Low energy behavior of random configurations}\label{AA}

\begin{figure}[h]
\includegraphics[width=\hsize]{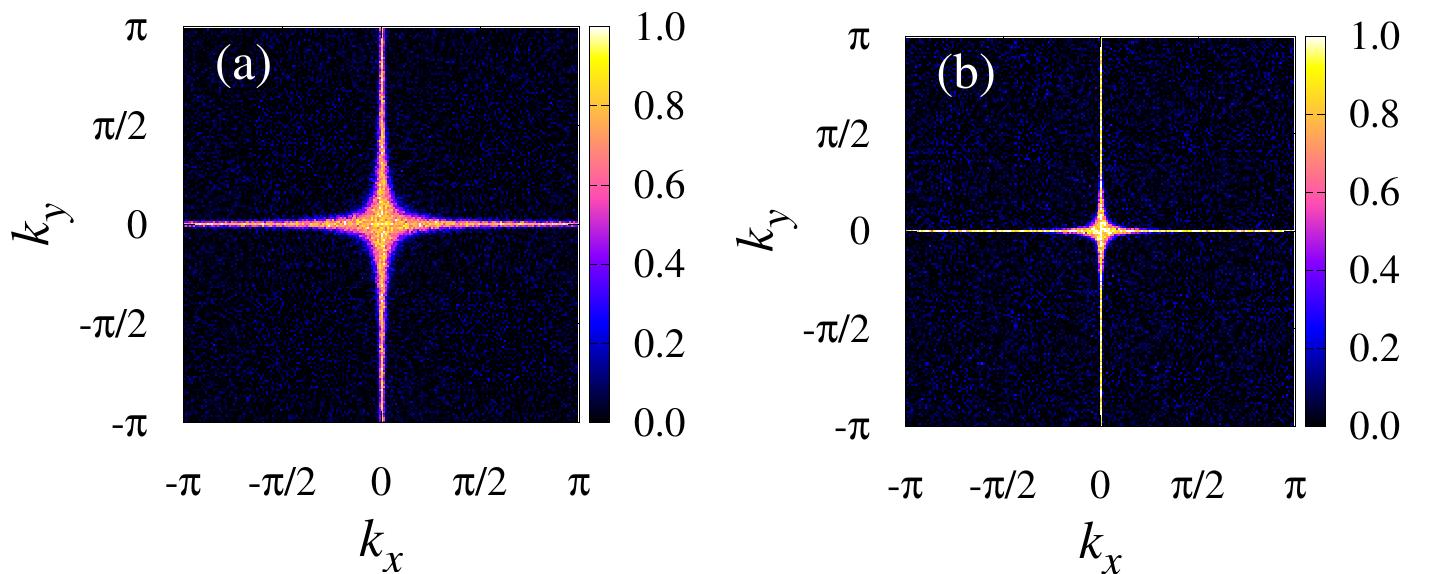}
\caption{$C(\vec{k},t)$ normalized by $C(\vec{0},t)$ for times 
$t=2^{12}$ (a) and $2^{16}$ (b).}
\label{figstrfct}
\end{figure}

Here we follow the analysis developed in Ref.~\cite{sala2021dynamics} for
a similar stochastic model. The authors consider a classical automaton on a
square lattice which also conserves the dipole moment similar to the model
we have considered. The microscopic dynamics is considered from the
perspective of the long time correlation function, defined as
$C(\vec{r},t)=\braket{n_{\vec{r}}(t)n_0(0)}$, where the expectation value
is over random initial conditions and realizations of stochastic
dynamics. The authors of Ref.~\cite{sala2021dynamics}  estimate the correlation function in
momentum space ($C(\vec{k},t)$) for a time much larger than the scale of
the microscopic dynamics (using both automaton dynamics and an effective analytical ansatz) and find sub-diffusive features with ``hidden''
modulated symmetries corresponding to certain patterns in the Brillouin zone  (see Fig.2b of Ref.~\cite{sala2021dynamics}).

In this appendix, we would like to check if this analysis can help identify the slow dynamics we observe for large scale modulated patterns. The energy
spectrum as a function of $\vec{k}$ has already been identified in a mean-field framework by Paramekanti \textit{et al.} in Ref.~\cite{paramekanti2002ring} for the form of ring-exchange used in our work, to be given by $E_{\vec{k}}\propto |\sin(k_x/2)\sin(k_y/2)|$. This trivially implies that the lines $k_x=0$ and $k_y=0$ host zero modes, and that this should be visible in $C(\vec{k},t)$ in the long time limit. Note that this profile does {\it not} explain the slow dynamics which we have explored in this work, as
that would at the minimum require the presence of slow modes along tilted
directions such as $k_x=k_y$. To check if this expectation is borne out by
 ring-exchange dynamics, we perform numerical evolutions of random
configurations of a $200\times 200$ lattice
for $t=2^{12}$ and $2^{16}$, shown in Fig.~\ref{figstrfct}.
We can clearly see that the profile is consistent with non-zero values
being present only along the lines $k_x=0$ and $k_y=0$ at long times. This
aspect of the dynamics is consistent with the mean-field
treatment developed in the body of the manuscript. This is to be expected since random configurations correspond to the $A\to 0$ limit of our study of modulated patterns, which was shown to be well-described by the ``vanilla'' mean-field dynamics.

\bibliography{slowdyn}

\end{document}